\documentclass[11pt]{article}
\usepackage{smile}

\usepackage{fullpage}
\usepackage{lscape}
\usepackage{bigints}
\usepackage{framed}
\usepackage{mdframed}
\usepackage{enumerate}
\usepackage[inline]{enumitem}
\usepackage[T1]{fontenc}
\usepackage{moresize}
\usepackage{bm}
\usepackage{bbm}
\usepackage{dsfont}
\usepackage{amsmath}
\usepackage{amssymb}
\usepackage{amsthm}
\usepackage{amsfonts}
\usepackage{stmaryrd}
\usepackage{array}
\usepackage{mathrsfs}
\usepackage{mathtools} 
\usepackage{extarrows}
\usepackage{stackrel}
\usepackage{relsize,exscale}
\usepackage{scalerel}
\usepackage[nodisplayskipstretch]{setspace}
\usepackage{color}
\usepackage[usenames,dvipsnames]{xcolor}
\usepackage{cancel}
\usepackage{soul}
\usepackage{undertilde}
\usepackage{xfrac}
\usepackage{siunitx}
\usepackage{graphicx}
\usepackage{float}
\usepackage{rotating}
\usepackage{subcaption}
\usepackage{overpic}
\usepackage[all]{xy}
\usepackage{tikz}
\usetikzlibrary{arrows,matrix,positioning,calc,automata,patterns}
\usepackage{booktabs}
\usepackage{dcolumn}
\usepackage{multirow}
\usepackage{diagbox}
\usepackage{tabularx}
\usepackage{verbatim}
\usepackage{listings}
\usepackage[ruled,vlined]{algorithm2e}
\usepackage{fancyvrb}
\usepackage{hyperref}
\usepackage[round]{natbib}
\usepackage{sectsty}
\usepackage{etoolbox}
\usepackage[capitalize,noabbrev]{cleveref}
\crefname{assumption}{Assumption}{Assumptions}
\newcounter{step}
\AtBeginEnvironment{proof}{\setcounter{step}{0}}
\crefname{step}{Step}{Steps}

\hypersetup{
    bookmarks=true,         
    unicode=false,          
    pdftoolbar=true,        
    pdfmenubar=true,        
    pdffitwindow=false,     
    pdfstartview={FitH},    
    pdftitle={My title},    
    pdfauthor={Author},     
    pdfsubject={Subject},   
    pdfcreator={Creator},   
    pdfproducer={Producer}, 
    pdfkeywords={key1, key2}, 
    pdfnewwindow=true,      
    colorlinks=true,        
    linkcolor=blue,         
    citecolor=blue,         
    filecolor=blue,         
    urlcolor=cyan           
}

\usepackage{stackengine}
\stackMath
\newcommand\tenq[2][1]{%
\def\useanchorwidth{T}%
\ifnum#1>1%
\stackunder[0pt]{\tenq[\numexpr#1-1\relax]{#2}}{\!\scriptscriptstyle\thicksim}%
\else%
\stackunder[1pt]{#2}{\!\scriptstyle\thicksim}%
\fi%
}

\makeatletter
\DeclareRobustCommand\widecheck[1]{{\mathpalette\@widecheck{#1}}}
\def\@widecheck#1#2{%
    \setbox\z@\hbox{\m@th$#1#2$}%
    \setbox\tw@\hbox{\m@th$#1%
       \widehat{%
          \vrule\@width\z@\@height\ht\z@
          \vrule\@height\z@\@width\wd\z@}$}%
    \dp\tw@-\ht\z@
    \@tempdima\ht\z@ \advance\@tempdima2\ht\tw@ \divide\@tempdima\thr@@
    \setbox\tw@\hbox{%
       \raise\@tempdima\hbox{\scalebox{1}[-1]{\lower\@tempdima\box
\tw@}}}%
    {\ooalign{\box\tw@ \cr \box\z@}}}
\makeatother

\newcommand{\X}{X}
\newcommand{\x}{x}
\newcommand{\sd}{\,\triangle\,}
\newcommand{\Er}{\mathrm{E}}
\newcommand{\dr}{\mathrm{d}}
\newcommand{\Ibbm}{\mathbbm{1}}
\newcommand{\Prb}{\mathrm{P}}

\newcommand{\as}{a.s.\ }

\newcommand{\lb}{\left( }
\newcommand{\rb}{\right) }

\newcommand{\lcb}{\left\{ }
\newcommand{\rcb}{\right\} }
\newcommand{\er}{e}

\newcommand{\n}{\mathbf{n}}
\newcommand{\Cc}{\mathcal{C}}
\newcommand{\Cf}{\mathfrak{C}}
\newcommand{\Lc}{\mathcal{L}}

\newcommand{\Kc}{\mathcal{K}}
\newcommand{\Ec}{\mathcal{E}}

\newcommand{\Sc}{\mathcal{S}}
\newcommand{\Zb}{\mathbb{Z}}
\newcommand{\Sb}{\mathbb{S}}
\newcommand{\Bf}{\mathfrak{B}}
\newcommand{\Af}{\mathfrak{A}}
\newcommand{\Df}{\mathfrak{D}}
\newcommand{\Ff}{\mathfrak{F}}
\newcommand{\Mf}{\mathfrak{M}}
\newcommand{\Gf}{\mathfrak{G}}
\newcommand{\pf}{\mathfrak{p}}

\newcommand{\Rb}{\mathbb{R}}

\newcommand{\Fc}{\mathcal{F}}
\newcommand{\Xc}{\mathcal{X}}
\newcommand{\Hc}{\mathcal{H}}
\newcommand{\Tr}{\mathrm{Tr}}
\newcommand{\var}{\mathrm{Var}}
\newcommand{\Diag}{\mathrm{Diag}}

\newcommand{\supp}{{\rm supp}}

\newcommand{\Step}[2][]{%
  \refstepcounter{step}%
  \paragraph*{Step \thestep: #2.}%
  \if\relax\detokenize{#1}\relax\else\label{#1}\fi
}

\numberwithin{equation}{section}

\newtheorem{theorem}{Theorem}[section]
\newtheorem{lemma}{Lemma}[section]

\newtheorem{assumption}{Assumption}[section]

\newtheorem{innercustomgeneric}{\customgenericname}
\providecommand{\customgenericname}{}
\newcommand{\newcustomtheorem}[2]{%
  \newenvironment{#1}[1]
  {%
   \renewcommand\customgenericname{#2}%
   \renewcommand\theinnercustomgeneric{##1}%
   \innercustomgeneric
  }
  {\endinnercustomgeneric}
}
\newcustomtheorem{customdefinition}{Definition}
\newcustomtheorem{customdefinitions}{Definitions}
\newcustomtheorem{customtheorem}{Theorem}
\newcustomtheorem{customassumption}{Assumption}
\newcustomtheorem{customlemma}{Lemma}
\newcustomtheorem{customexample}{Example}
\theoremstyle{definition}

\newtheorem{remark}{Remark}[section]

\newlist{assparts}{enumerate}{1}
\setlist[assparts,1]{label=(\roman*), ref=\theassumption-(\roman*), itemsep=0pt}

\makeatletter
\newcommand{\mylabel}[2]{#2\def\@currentlabel{#2}\label{#1}}
\makeatother

\crefname{asspartsi}{Assumption}{Assumptions}

\graphicspath{{./fig3/}}

\allowdisplaybreaks

\begin{document}

\setlength{\abovedisplayskip}{5pt}
\setlength{\belowdisplayskip}{5pt}
\setlength{\abovedisplayshortskip}{5pt}
\setlength{\belowdisplayshortskip}{5pt}
\hypersetup{colorlinks,breaklinks,urlcolor=blue,linkcolor=blue}

\title{\LARGE Bias correction for Chatterjee's graph-based correlation coefficient}

\author{Mona Azadkia\thanks{Department of Statistics, London School of Economics and Political Science; {\tt m.azadkia@lse.ac.uk}},~~~
	    Leihao Chen\thanks{Korteweg-de Vries Institute for Mathematics, University of Amsterdam, Netherlands; {\tt l.chen2@uva.nl}}, ~~and~~
	    Fang Han\thanks{Department of Statistics, University of Washington, Seattle, WA 98195, USA; {\tt fanghan@uw.edu}}
}

\date{\today}

\maketitle

\vspace{-1em}

\begin{abstract}
\cite{Azadkia21simple} recently introduced a simple nearest neighbor (NN) graph-based correlation coefficient that consistently detects both independence and functional dependence. Specifically, it approximates a measure of dependence that equals 0 if and only if the variables are independent, and 1 if and only if they are functionally dependent. However, this NN estimator includes a bias term that may vanish at a rate slower than root-$n$, preventing root-$n$ consistency in general. In this article, we (i) analyze this bias term closely and show that it could become asymptotically negligible when the dimension is smaller than four; and (ii) propose a bias-correction procedure for more general settings. In both regimes, we obtain estimators (either the original or the bias-corrected version) that are root-$n$ consistent and asymptotically normal.

\end{abstract}


\section{Introduction}

Sourav Chatterjee, in his groundbreaking paper \citep{chatterjee2020new}, introduced a novel and elegant approach to estimating the dependence measure of \cite{MR3024030}. This measure captures both independence and perfect functional dependence: it equals 0 if and only if the two random scalars are independent, and 1 if and only if one is a measurable function of the other almost surely. Chatterjee’s work led to a new rank-based statistic, now widely known as Chatterjee’s rank correlation.

Among the notable extensions of this work is the contribution by \cite{Azadkia21simple}, who generalized Chatterjee’s rank correlation to multivariate settings through a sophisticated integration of rank statistics and nearest neighbor (NN) graphs. However, unlike Chatterjee’s original rank correlation—which has been shown to be root-$n$ consistent and asymptotically normal under broad conditions \citep{lin2022limit,kroll2024asymptotic}—the NN-based correlation coefficient of \cite{Azadkia21simple} suffers from a bias term that may decay at a rate slower than root-$n$, rendering the estimator not root-$n$ consistent in general.

This article examines the bias of the NN-based correlation coefficient in greater depth and contributes in two main directions. First, building on the thought-provoking work of \cite{viel2025convergenceratenearestneighbour}, which analyzed the bias of another NN-based estimator in causal inference, we show that the bias of the NN-based correlation coefficient could become asymptotically negligible when the dimension $d \le 3$. This finding advances the current understanding, which establishes analogous asymptotic negligibility only in the univariate case $d=1$.

Second, for more general scenarios in which the bias may no longer be ignorable, we propose a regression-adjustment technique and derive the large-sample properties of the resulting bias-corrected estimator of Chatterjee’s NN graph-based correlation coefficient. Our approach is inspired, once again, by related developments in causal inference, including the influential ideas of \cite{rubin1973use} and the theoretical foundations laid out in \cite{abadie2011bias}, \cite{lin2021estimation}, \cite{lin2022regression}, and \cite{cattaneo2024rosenbaum}. Leveraging these insights, we develop a new theory for nonparametric ridge regression and demonstrate that the proposed correction can asymptotically eliminate the bias of the original NN estimator, thereby restoring root-$n$ consistency without affecting the limiting variance.

In both directions, the original NN-based estimator of \cite{Azadkia21simple} (when $d \le 3$) and the newly proposed bias-corrected estimator (for more general settings) share the same attractive property: each achieves root-$n$ consistency and asymptotic normality with an identical limiting variance, regardless of the dimensionality of the covariates. This robustness allows seamless integration with both analytical variance estimators \citep{lin2022limit} and bootstrap-based procedures \citep{dette2025simple}, thereby facilitating valid inference.


\vspace{0.2cm}

{\bf Notation.} In the following, for any two real numbers $a, b$, we write $a \wedge b := \min\{a,b\}$ and $a \vee b := \max\{a,b\}$. For any two real sequences $\{a_n\}_{n\geq 1}$ and $\{b_n\}_{n\geq 1}$, we denote $a_n= O(b_n)$ (or equivalently, $a_n\lesssim b_n$) if there exists a constant $C>0$ such that $|a_n|\le C|b_n|$ for all sufficiently large $n$, and $a_n= o(b_n)$ if $|a_n|/|b_n|\to 0$ as $n\to \infty$. For any two real sequences $\{a_n\}, \{b_n\}$, we write $a_n \asymp b_n$ if both $a_n\lesssim b_n$ and $b_n\lesssim a_n$ hold true. We write $O_{\Prb}$ and $o_{\Prb}$  if $O(\cdot)$ and $o(\cdot)$ hold in probability, respectively.

\section{Chatterjee's NN graph-based correlation coefficient}

We begin with a brief review of the current understanding of \cite{Azadkia21simple}'s NN graph-based correlation coefficient. Consider $(X,Y) \in \Rb^d \times \Rb$ to be a pair of random variables. To quantify the strength of dependence between $X$ and $Y$, \cite{MR3024030} introduced the following population quantity, now referred to as the \textit{Dette–Siburg–Stoimenov dependence measure}:
\[
    T = T(Y,X) := \frac{\int \var\big(\Er[\Ibbm(Y \geq t) \mid X]\big) \dr \mu(t)}{\int \var\big(\Ibbm(Y \geq t)\big) \dr \mu(t)},
\]
where $\mu$ denotes the law of $Y$ and $\Ibbm(\cdot)$ is the indicator function. As shown in \cite{MR3024030} and further developed in \cite{Azadkia21simple}, $T$ satisfies the desirable properties of being
\begin{enumerate}[itemsep=-.5ex,label=(\roman*)]
\item $T = 0$ if and only if $Y$ is independent of $X$;
\item $T = 1$ if and only if $Y$ is a measurable function of $X$ almost surely.
\end{enumerate}

Let $(X_1,Y_1),\ldots,(X_n, Y_n)$ be the sample. For each $i \in [n] := \{1,2,\ldots,n\}$, define
\[
R_i := \sum_{j=1}^n \Ibbm(Y_j \leq Y_i), \quad \text{and} \quad N_i := \argmin_{j \ne i} \|X_i - X_j\|,
\]
where $R_i$ is the rank of $Y_i$ among $\{Y_1, \ldots, Y_n\}$ and $N_i$ indexes the nearest neighbor (NN) of $X_i$ under the Euclidean metric, $\|\cdot\|$. To estimate $T$ based on the sample $\{(X_i, Y_i)\}_{i=1}^n$, \cite{Azadkia21simple} proposed the following NN-based rank correlation coefficient:
\[
\hat T_n = \hat{T}_n(Y,X) := \frac{6}{n^2 - 1} \sum_{i=1}^n \min\big\{R_i, R_{N(i)}\big\} - \frac{2n + 1}{n - 1}.
\]

The asymptotic properties of $\hat T_n$ are summarized in the following theorem under the following assumption of the data generating scheme.

\begin{assumption}\label{asump:dgp}
$(X,Y), (X_1,Y_1),\ldots, (X_n,Y_n)$ are independently sampled from a fixed and continuous cumulative distribution function (CDF) $F_{X,Y}$.
\end{assumption}

\begin{theorem}\label{thm:basic}
Assume Assumption \ref{asump:dgp}.
\begin{enumerate}[itemsep=-.5ex,label=(\roman*)]
\item \citet[Theorem 2.2]{Azadkia21simple}: $\hat T_n$ converges almost surely to $T$.
\item \cite{deb2020kernel} and \citet[Theorem 3.1]{shi2021ac}: under the assumption that $X$ and $Y$ are independent and $X$ is absolutely continuous, $n^{1/2} \hat T_n$ converges in distribution to a normal distribution with mean zero and a distribution-free variance.
\item \citet[Theorem 1.1]{lin2022limit}: under the assumption that $Y$ is not a measurable function of $X$ almost surely, we have
\[
\frac{\hat T_n - \Er[\hat T_n]}{\sqrt{\var[\hat T_n]}} \longrightarrow N(0,1) \quad \text{in distribution}.
\]
\item\label{thm:basic-3} \citet[Theorem 4.1]{Azadkia21simple} and \citet[Proposition 1.1]{lin2022limit}: suppose there exist constants $\beta, C, C_1, C_2 > 0$ such that for all $t \in \bR$ and $x, x' \in \bR^d$,
\begin{align*}
    &\Big| \Prb(Y \ge t \mid X = x) - \Prb(Y \ge t \mid X = x') \Big| \le C\big(1 + \|x\|^\beta + \|x'\|^\beta\big)\|x - x'\|, \\
    \text{and} \quad &\Prb(\|X\| \ge t) \le C_1 e^{-C_2 t},
\end{align*}
then
\[
    \Er[\hat T_n] - T = O\left( \frac{(\log n)^{d + \beta + 1 + \ind(d=1)}}{n^{1/d}} \right).
\]
\item \citet[Proposition 1.2]{lin2022limit}: the limit of $n \var(\hat T_n)$ exists, is finite, and equals zero if and only if $Y$ is a measurable function of $X$ almost surely.
\item \citet[Theorem 1.2]{lin2022limit} and \cite{dette2025simple}: consistent estimators of the limiting variance $n \var(\hat T_n)$ exist.
\end{enumerate}
\end{theorem}

Theorem~\ref{thm:basic} characterizes the asymptotic behavior of $\hat T_n$. In particular, Part~\ref{thm:basic-3} highlights the main obstacle preventing $\hat T_n$ from achieving root-$n$ consistency: the presence of a theoretically non-negligible bias term. This is not unexpected for researchers familiar with NN-based statistics. Indeed, bias correction has played a central role in related areas such as NN graph-based estimation for causal inference \citep{abadie2011bias, lin2021estimation} and entropy estimation \citep{MR3909934}. In the next section, we will present our approach to addressing this final gap for $\hat T_n$.

Beyond the above theoretical results, it is worth acknowledging ongoing efforts in this research direction, including deeper investigations of $T$ \citep{bucher2024lack}, further analyses of $\hat T_n$ \citep{shi2020power, bickel2022measures, lin2024failure, han2024azadkia, auddy2021exact, zhang2024asymptotic}, exploration of new dependence measures satisfying the defining properties of $T$ \citep{griessenberger2022multivariate, ansari2025directextensionazadkia, strothmann2024rearranged, azadkia2025new}, and extensions of $\hat T_n$-type ideas to broader statistical contexts \citep{huang2020kernel, azadkia2021fast, gamboa2022global, ansari2023quantifying, gao2024family, fuchs2024quantifying,  tran2024rank, zhang2025extensions}. More recent developments include \cite{zhou2025association} and \cite{olivares2025powerful}. For broader overviews and related perspectives, we refer the reader to the surveys conducted by \cite{han2021extensions} and \cite{chatterjee2022survey}.

\section{Bias analysis}\label{sec:bias_analysis}

We begin with a closer analysis of the bias of $\hat T_n$. To this end, we first recall the closed-form expression for the bias term $\Er[\hat T_n]-T$, as derived in \cite{lin2022limit}.

\begin{lemma}[Equation (3.1), \cite{lin2022limit}]\label{lem:bias}
Assume Assumption \ref{asump:dgp}. Then
\begin{align}\label{eq:bias}
\Er[\hat T_n] - T =&\ 
6\left\{
\Er\left[\Ibbm\left(Y_2 \le Y_1 \wedge Y_{N(1)}\right)\right]
- \Er\left[\Ibbm\left(Y_2 \le Y_1 \wedge \tilde{Y}_1\right)\right]
\right\} \notag\\
&- \frac{6}{n+1}\,\Er\left[\Ibbm\left(Y_2 \le Y_1 \wedge Y_{N(1)}\right)\right]
+ \frac{6n}{n^2-1}\,\Er\left[\Ibbm\left(Y_1 \le Y_1 \wedge Y_{N(1)}\right)\right]
- \frac{3}{n-1},
\end{align}
where $\tilde Y_1$ is an independent sample from the conditional distribution of $Y$ given $X=X_1$.
\end{lemma}

Define the leading bias term in \eqref{eq:bias} by
\[
L \coloneqq 
\Er\Big[\Ibbm\!\big(Y_2 \le Y_1 \wedge Y_{N(1)}\big)\Big]
-
\Er\Big[\Ibbm\!\big(Y_2 \le Y_1 \wedge \tilde Y_1\big)\Big].
\]
Lemma~\ref{lem:bias} then directly yields
\[
\Er[\hat T_n] - T = 6L + O(n^{-1}).
\]
Thus, analyzing the bias reduces to understanding the behavior of $L$.

To this end, introduce the conditional mean function
\[
G_{\x}(t) \coloneqq \Er\big[\Ibbm(Y \ge t)\mid X=\x\big].
\]
A simple calculation shows that
\[
L
= \Er\!\Big[F_Y\!\big(Y_1 \wedge Y_{N(1)}\big) - F_Y\!\big(Y_1 \wedge \tilde Y_1\big)\Big]
= \int \Er\!\left[G_{X_1}(t)\,G_{X_{N(1)}}(t) - G_{X_1}^2(t)\right] \dr\mu(t),
\]
highlighting the central role played by the conditional mean function $G_x(\cdot)$ in the analysis of $\Er[\hat T_n]-T$. This is a key insight already emphasized in the original paper \citep{Azadkia21simple}.

Equation~\eqref{eq:bias} is a classical representation of the bias of NN-based functional estimators. As shown in \cite{Azadkia21simple} (cf.\ Theorem~\ref{thm:basic}\ref{thm:basic-3}), a direct analysis of \eqref{eq:bias} implies that the resulting estimator converges at a rate slower than $n^{-1/2}$ whenever $d \ge 2$. In the broader literature, \cite{singh2016finite} were the first to suggest that such functional NN estimators may nevertheless achieve parametric convergence rates in multivariate settings, provided the underlying density is sufficiently smooth, even though the NN density estimator itself generally does not adapt automatically to such smoothness \citep{zhao2022analysis}. Subsequently, and more closely related to our work, \cite{viel2025convergenceratenearestneighbour} analyzed the bias of an NN-based estimator in causal inference and demonstrated that the phenomenon identified by \cite{singh2016finite} extends to this class of NN estimators as well.

In this section, we follow the argument of \cite{viel2025convergenceratenearestneighbour} and show that the bias term in \eqref{eq:bias} can also be asymptotically negligible. To prepare for the analysis, we introduce some additional notation. For a twice continuously differentiable function
\[
f: U \to \Rb, \qquad U \subseteq \Rb^d \ \text{open},
\]
we denote its gradient and Hessian at $x$ by $\nabla f(x)$ and $\nabla^{2} f(x)$, respectively. Let $\Xc \coloneqq \supp(X)$ denote the support of $X$, and define
\[
\delta(x) \coloneqq 
\begin{cases}
\mathrm{dist}(x,\partial \Xc), & \text{if } x \in \Xc,\\[3pt]
-\mathrm{dist}(x,\partial \Xc), & \text{if } x \notin \Xc,
\end{cases}
\]
where
\[
\mathrm{dist}(x,\partial \Xc)
= \inf_{z \in \partial \Xc} \|x - z\|
\]
denotes the Euclidean distance from $x$ to the boundary $\partial \Xc$ of~$\Xc$. Finally, for any positive integer $b$, we use $\Hc^b$ to denote the $b$-dimensional Hausdorff measure.

The following assumption regulates the behavior of $X$ and imposes smoothness conditions on the conditional mean function $G_x(t)$ with respect to $x$. Similar conditions appear in \citet[Assumptions~X5, X6-1, X6-2]{viel2025convergenceratenearestneighbour}.

\begin{assumption}\label{assum:ass_4}
Let $\Xc \coloneqq \supp(X) \subseteq \Rb^d$. We assume: 
\begin{assparts}[label=(\roman*),ref=\theassumption-(\roman*),itemsep=0pt]

\item \label{ass:G_function}
For every $t \in \supp(\mu)$, there exists a twice continuously differentiable function 
\[
\tilde G_{\cdot}(t): \Rb^d \to [0,1]
\]
such that $\tilde G_{\cdot}(t) = G_{\cdot}(t)$ on $\Xc$, and
\[
\sup_{t \in \supp(\mu),\, x' \in \Rb^d}
\Big( \|\nabla_x \tilde G_{x'}(t)\| + \|\nabla_x^2 \tilde G_{x'}(t)\|_2 \Big) < \infty,
\]
where $\|\cdot\|_2$ denotes the matrix spectral norm.

\item \label{ass:density}
$X$ admits a Lebesgue density $f_X$ satisfying 
\[
\inf_{x \in \Xc} f_X(x) \ge \pf > 0
\]
for some constant $\pf$, and $f_X$ is Lipschitz continuous on $\Xc$.

\item \label{ass:compact}
$\Xc$ is compact.

\item \label{ass:Lip_boundary}
$\Xc$ has a Lipschitz boundary \cite[Definition~1.2.1.1]{Grisvard11Elliptic_nonsmooth_domains}.\footnote{Note that if $\emptyset \ne\mathring{\Xc}$ (the interior of $\Xc$) is bounded and convex, then $\Xc$ has a Lipschitz boundary; cf.\ \citet[Corollary~1.2.2.3]{Grisvard11Elliptic_nonsmooth_domains}.} 
\end{assparts}
\end{assumption}

Assumption~\ref{assum:ass_4} yields the following result, which establishes the asymptotic negligibility of the bias even when $d>1$.

\begin{theorem}[Bias rate]\label{thm:bias_rate}
    Assume Assumptions~\ref{asump:dgp} and \ref{assum:ass_4}. Then, as $n \to \infty$,
    \[
        \Er[\hat{T}_n] - T = O\big(n^{-2/d} + n^{-1}\big).
    \]
    Consequently, whenever $d \le 3$,
    \begin{align}\label{eq:clt}
        n^{1/2}(\hat T_n - T) \;\to\; N(0,\sigma_T^2) 
        \quad \text{in distribution},
    \end{align}
    where $\sigma_T^2$ is the asymptotic variance of $\hat T_n$ as given in \citet[Proposition~1.2(i)]{lin2022limit}.
\end{theorem}

Theorem~\ref{thm:bias_rate} provides a convergence rate for $\Er[\hat{T}_n]-T$ that is typically sufficient for establishing the central limit theorem \eqref{eq:clt}. In particular, it shows that—under the stated conditions—the bias is $o(n^{-1/2})$ whenever $d \le 3$. Under additional smoothness assumptions, however, one can derive a more refined expansion of the bias and show that the above rate is, in general, unimprovable and therefore sharp. Such refinements follow from a higher-order expansion of the bias term, paralleling the arguments of \citet[Theorem~2]{viel2025convergenceratenearestneighbour}.

\begin{assumption}\label{assum:ass_5}
We assume the following is true.

\begin{assparts}[label=(\roman*),ref=\theassumption-(\roman*),itemsep=0pt]

    \item There exists a continuously differentiable function 
    \[
        \tilde f_X : \Rb^d \to [0,\infty)
    \]
    such that $\tilde f_X(x) = f_X(x)$ for all $x \in \Xc$, and $\tilde f_X$ is $C^{1}$ on $\Rb^d$. Moreover, there exists $\alpha \in (0,1)$ such that
    \[
        \|\nabla \tilde f_X\|_{C^{0,\alpha}(\Xc)}
        \coloneqq 
        \sup_{x \neq y \in \Xc}
        \frac{\|\nabla \tilde f_X(x) - \nabla \tilde f_X(y)\|}{\|x - y\|^{\alpha}}
        < \infty.
    \]

    \item There exists $\beta \in (0,1)$ such that the function $\tilde G$ in Assumption~\ref{ass:G_function} further satisfies
    \[
    \sup_{t \in \supp(\mu)}
    \Big(
        \sup_{x'} \|\nabla_x \tilde G_{x'}(t)\|
        + \sup_{x'} \|\nabla_x^{2} \tilde G_{x'}(t)\|_2
        + \|\nabla_x^{2} \tilde G_{\cdot}(t)\|_{C^{0,\beta}(\Xc)}
    \Big)
    < \infty.
    \]

    \item \label{ass:C2_boundary} $\Xc$ has a $C^{2}$ boundary \cite[Definition~1.2.1.1]{Grisvard11Elliptic_nonsmooth_domains}. 
\end{assparts}
\end{assumption}

\begin{theorem}[Bias expansion]\label{thm:bias_expansion}
   Assume Assumptions \ref{asump:dgp}, \ref{assum:ass_4}, and \ref{assum:ass_5}. We then have, as $n\to \infty$,
    \[
        \Er[\hat{T}_n]-T=(\Cf_1+\Cf_3) n^{-\frac{2}{d}}+\Cf_2 n^{-1}+o(n^{-\frac{2}{d}}+n^{-1}),
    \]
    where
    \[
        \begin{aligned}
            \Cf_1&:=-6\Gamma(1+2/d)v_d^{-\frac{2}{d}}\int_\Xc f_X(x)^{1-\frac{2}{d}}d^{-1}\big(\nabla \log f_X(x)\cdot \Af(x)+\frac{1}{2}\Tr(\Bf(x))\big)\dr x,\\
            \Af(x)&:=\int G_x(t)\nabla_x G_x(t)\dr \mu(t), \quad \quad \Bf(x):=\int G_x(t)\nabla_x^2 G_x(t)\dr \mu(t),\\
            \Cf_2&:=-6\int \Er\big[G_X(t)^2\big]\dr \mu(t),\\
            \Cf_3&:=-3(2^{2/d}-1)v_d^{-2/d}\Gamma(1+2/d)\int_{\partial \Xc} f_X(y)^{1-2/d}\Gf(y)\dr \Hc^{d-1}(y),\\
            \Gf(x)&:=\int \big(\nabla\delta(x)\cdot \nabla_xG_x(z)\big)G_x(z)\dr \mu(z),
        \end{aligned}
    \]
    where $v_d$ denotes the volume of Euclidean unit ball in $\Rb^d$ and $\Tr(\cdot)$ outputs the trace of the input matrix.
\end{theorem}

\section{Bias correction}

\subsection{Method}

Section~\ref{sec:bias_analysis} outlines the landscape of the bias of $\hat T_n$ and provides both positive and negative insights. On the positive side, Theorem~\ref{thm:bias_rate} shows that the bias is asymptotically negligible whenever $d \le 3$. On the other hand, Theorem~\ref{thm:bias_expansion} demonstrates that, in general, the bias cannot converge at a rate faster than that established in Theorem~\ref{thm:bias_rate}. Consequently, a bias correction becomes necessary whenever $d \ge 4$.

As a matter of fact, Lemma~\ref{lem:bias} naturally motivates a bias-corrected version of $\hat T_n$. Recall that
\[
\Er[\hat T_n] - T = 6L + O(n^{-1}).
\]
To obtain a root-$n$ consistent estimator of $T$, it suffices to construct an estimator $\hat L_n$ such that 
\[
\hat L_n - L = o_{\Prb}(n^{-1/2}),
\]
in which case
\[
n^{1/2}(\hat T_n - 6\hat L_n - T)
    = n^{1/2}(\hat T_n - \Er[\hat T_n]) + o_{\Prb}(1).
\]

Toward this goal, recall that
\[
L 
= \Er\!\left[F_Y\!\big(Y_1 \wedge Y_{N(1)}\big) - F_Y\!\big(Y_1 \wedge \tilde Y_1\big)\right]
= \int \Er\!\left[G_{X_1}(t)\, G_{X_{N(1)}}(t) - G_{X_1}^2(t)\right] \dr \mu(t),
\]
suggesting that a natural estimator of $L$ may be obtained by replacing the integral, expectations, and the conditional mean function $G_x(\cdot)$ with their empirical counterparts.

The procedure then proceeds as follows.
\begin{enumerate}[itemsep=-.5ex,label=(\roman*)]
\item We use the full data to estimate the {\it bivariate regression function} $G_x(\cdot)$, yielding an estimator $\hat G_x(\cdot)$, and to approximate the expectation
\[
\Er\left[\hat G_{X_1}(t)\, \hat G_{X_{N(1)}}(t) - \left\{\hat G_{X_1}(t)\right\}^2\right]
\]
via the average
\[
\hat \Er \left[\hat G_{X_1}(t)\, \hat G_{X_{N(1)}}(t) - \left\{\hat G_{X_1}(t)\right\}^2\right]
:= \frac{1}{n} \sum_{i=1}^n \left[
\hat G_{X_i}(t)\, \hat G_{X_{N(i)}}(t)
- \left\{\hat G_{X_i}(t)\right\}^2
\right].
\]
\item We next use the empirical distribution of $Y_1,\ldots,Y_n$ to approximate the integral over $\mu$, yielding the following bias estimator that takes a natural U-statistic form:
\begin{align*}
\hat L_n :=& \frac{1}{n(n-1)} \sum_{i\ne j\in[n]}
\left[\hat G_{X_i}(Y_j)\, \hat G_{X_{N(i)}}(Y_j)
- \left\{\hat G_{X_i}(Y_j)\right\}^2\right].
\end{align*}
\end{enumerate}

The resulting {\it bias-corrected} NN graph-based correlation coefficient is then given by
\[
\hat T_n^{\mathrm{bc}} := \hat T_n - 6 \hat L_n. 
\]

\subsection{Theory}\label{sec:theory}

Before stating the main result of this section, we introduce additional notation. For any function $f: \Rb^d \to \Rb$, define its $L^{\infty}$ norm over the support of $X$ as
\[
\norm{f}_{\infty} := \sup_{x \in \supp(F_X)} |f(x)|,
\]
where $F_X$ denotes the CDF of $X$ and $\supp(\cdot)$ denotes the support of a distribution or law. For any positive integer $q$, let $\Lambda_q$ denote the set of multi-indices of total degree $q$:
\[
\Lambda_q := \left\{ \alpha = (\alpha_1, \ldots, \alpha_d) : \sum_{i=1}^d \alpha_i = q,\ \alpha_i \in \Zb^{\geq 0}~~\text{for all }i \in [d] \right\},
\]
where $\Zb^{\geq 0}$ is the set of nonnegative integers. For each $\alpha \in \Lambda_q$, we adopt the following multi-index notation for derivatives:
\[
D^\alpha f = \frac{\partial^{|\alpha|} f}{\partial \x^\alpha} = \frac{\partial^{\alpha_1 + \cdots + \alpha_d} f}{\partial x_1^{\alpha_1} \cdots \partial x_d^{\alpha_d}},\quad
|\alpha| = \sum_{i=1}^d \alpha_i,\quad
\alpha! = \prod_{i=1}^d \alpha_i!,\quad
\x^\alpha = \prod_{i=1}^d x_i^{\alpha_i}.
\] 
For any positive integer $K$, let $I_K$ denote the identity matrix of dimension $K$. Lastly, for any nonnegative random variable $Z$ and any $s>0$, we denote
\[
\Er^{s}[Z] := (\Er[Z])^{s}.
\]

\subsubsection{A general theory for bias correction}

To facilitate presentation, we begin by introducing general conditions for the bias-corrected estimator $\hat T_n^{\rm bc}$ under certain high-level assumptions on the estimator of $G_x(\cdot)$. These assumptions will be verified in Section \ref{sec:rls} for a specific class of ridge least squares estimators.


We first impose regularity conditions on $F_X$.

\begin{assumption}[Regularity conditions for $F_X$]\label{assum:x} 
The distribution of $X$ satisfies that 
\begin{enumerate}[itemsep=0pt,label=(\roman*)]    
\item\label{assum:x1} its density is bounded away from 0 in its support;
\item\label{assum:x2}  for any $r > 0$ and all sufficiently large $n$,
\[
\Er^{1/r} \|X_1 - X_{N(1)}\|^r \leq C_r n^{-1/d},
\]
where $C_r$ is a constant only depending on $r$.
\end{enumerate}
\end{assumption}
Assumption \ref{assum:x}\ref{assum:x1} is standard in the NN literature. On the other hand, we refer the readers of interest to \cite{evans2002asymptotic} for distributions that satisfy Assumption \ref{assum:x}\ref{assum:x2}.

Next, we assume sufficient smoothness of the function $G_{\cdot}(t)$, requiring a higher degree than \citet[Assumption (A1)]{Azadkia21simple}, but similar to what was posed in \citet[Assumption 4.4(iii)]{lin2021estimation} and \citet[Assumption 5(iii)]{cattaneo2024rosenbaum}.

\begin{assumption}[Smoothness conditions for $G_x(t)$]\label{assum:distribution} 
The followings hold:
\begin{enumerate}[itemsep=0pt,label=(\roman*)]    
\item $\max_{\alpha\in\Lambda_1} \|D^\alpha G_{\cdot}(t)\|_{\infty}$ is uniformly bounded over $t \in \supp(\mu)$;
\item $\max_{\alpha\in \Lambda_{\lfloor d/2 \rfloor +1}} \|D^\alpha G_{\cdot}(t)\|_{\infty}$ is uniformly bounded over $t \in \supp(\mu)$.
\end{enumerate}
\end{assumption}

Finally, we assume regularity of the estimator $\hat G_{\cdot}(t)$, in line with \citet[Assumption 4.5]{lin2021estimation} and \citet[Assumption 6]{cattaneo2024rosenbaum}.

\begin{assumption}[Regularity conditions for $\hat{G}_x(t)$]\label{assum:estimator}
The followings hold:
\begin{enumerate}[itemsep=0pt,label=(\roman*)]   
\item the integral
\[
\int \Er^{1/2}\left[\max_{\alpha \in \Lambda_{\lfloor d/2 \rfloor +1}} \Big\|D^\alpha \hat{G}_{\cdot}(t)\Big\|_{\infty}^2 \right] \dr \mu(t) = O(1);
\] 
\item for each $\ell \in \{0, \ldots, \lfloor d/2 \rfloor\}$, there exists $\gamma_\ell > \max\left\{ \frac{1}{2} - \frac{\ell \vee 1}{d}, 0 \right\}$ such that 
\[
\int \Er^{1/2}\left[\max_{\alpha\in\Lambda_\ell} \Big\|D^\alpha \hat{G}_{\cdot}(t) - D^\alpha G_{\cdot}(t)\Big\|_{\infty}^2\right] \dr \mu(t) = O(n^{-\gamma_\ell});
\]
\item the integral 
\[
\int\Er\Big[\Big\|\hat G_{\cdot}(t)-G_{\cdot}(t)\Big\|_{\infty}^2\Big]
        \,\dr\mu(t)= o\bigl(n^{-1/2}\bigr).
\]
\end{enumerate}
\end{assumption}

\begin{theorem}[Main result]\label{thm:convergence_B}
Assume Assumption \ref{asump:dgp} and Assumptions \ref{assum:x}-\ref{assum:estimator}. Then the followings hold:
\begin{enumerate}[itemsep=0pt,label=(\roman*)]      
\item $n^{1/2}\Er\big|\hat L_n - L\big| \to 0$;
\item $n^{1/2}(\hat T_n^{\rm bc} - T) \to N(0, \sigma^2)$ in distribution, where $\sigma^2$ is the asymptotic variance of $\hat T_n$ as in \citet[Proposition 1.2(i)]{lin2022limit}.
\end{enumerate}    
\end{theorem}

\begin{remark}
It is worth comparing the assumptions made in this section with those imposed in Section~\ref{sec:bias_analysis}. Recall that Assumptions~\ref{assum:ass_4} and \ref{assum:ass_5} place smoothness requirements on the density $f_X$ and the function $G$. In fact, when $d \le 3$, Assumption~\ref{ass:G_function} already implies Assumption~\ref{assum:distribution}, while Assumptions~\ref{ass:density}, \ref{ass:compact}, and \ref{ass:Lip_boundary} together imply Assumption~\ref{assum:x} (cf.\ Lemma~\ref{lem:Lip_boundary_interior_nondegeneracy} ahead). Thus, the bias-correction approach generally requires weaker smoothness assumptions, albeit at the cost of introducing additional steps to correct the bias.
\end{remark}

\subsubsection{Ridge least squares}\label{sec:rls}

The remaining task is to verify Assumption \ref{assum:estimator}, which cannot be directly addressed using standard nonparametric regression results \citep{newey1997convergence, chen2018optimal, belloni2015some}. In particular, unlike previous work, our setting requires control of estimation error in expectation rather than in probability, necessitating a more stable estimator.

To this end, we employ ridge regularization via \emph{ridge least squares} \citep{tuo2024asymptotic, kurisu2024series}. Following the setup of \citet[Section 4]{cattaneo2024rosenbaum}, let
\[
p_K(\cdot) = (p_{1K}(\cdot), \ldots, p_{KK}(\cdot))^\top \in \Rb^K
\]
be a $K$-dimensional vector of basis functions capable of approximating 
\[
\psi_t(\cdot) := G_{\cdot}(t), \quad \text{for any } t \in \supp(\mu).
\]
We estimate
\[
G_x(t) = \Er[\Ibbm(Y \ge t) \mid X = x]
\]
by projecting $\Ibbm(Y \ge t)$ onto the span of the basis functions. This corresponds to nonparametric linear probability models.

Define
\[
\psi_{t,K}(\x) := p_K(\x)^\top \beta_{t,K}, \quad \hat{\psi}_{t,K}(\x) := p_K(\x)^\top \hat{\beta}_{t,K},
\]
where $\beta_{t,K}$ is the population $L^2$ projection:
\[
\beta_{t,K} := \argmin_{b \in \Rb^K} \Er\left[(\psi_t(\X_1) - p_K(\X_1)^\top b)^2\right],
\]
and $\hat\beta_{t,K}$ is the ridge estimate:
\[
\hat{\beta}_{t,K} := \argmin_{b \in \Rb^K} \left\{ \frac{1}{n} \sum_{i=1}^n (\Ibbm(Y_i \ge t) - p_K(\X_i)^\top b)^2 + \lambda_n \|b\|^2 \right\},
\]
with regularization parameter $\lambda_n > 0$. Introduce further
  \begin{align*}
    &P \coloneqq
      \bigl(p_K(\X_1),\ldots,p_K(\X_n)\bigr)^\top
      \in \Rb^{n\times K},\qquad \Ibbm(Y_{[n]}\ge t)\coloneqq
      \bigl(\Ibbm(Y_1\ge t),\ldots,\Ibbm(Y_n\ge t)\bigr)^\top \in \Rb^n,\\
    \text{and}~~~&Q \coloneqq \Er\bigl[p_K(\X)\,p_K(\X)^\top\bigr]\in \Rb^{K\times K}.
  \end{align*}
  
The following quantities play a key role in characterizing the behavior of series estimators and their associated approximation errors:
\[
\underline{\lambda}_K := \lambda_{\min}(Q), \quad \zeta_{r,K} := \max_{\alpha \in \Lambda_r} \sup_{\x} \|D^\alpha p_K(\x)\|, \quad \vartheta_{r,K}^t := \max_{\alpha \in \Lambda_r} \|D^\alpha \psi_t - D^\alpha \psi_{t,K}\|_{\infty}.
\]
Here, \(\lambda_{\min}(Q)\) denotes the smallest eigenvalue of \(Q\); \(\zeta_{r,K}\) measures the smoothness of the basis functions \(p_K\); and \(\vartheta_{r,K}^t\) captures the best possible order-\(r\) approximation error of the basis \(p_K\) in representing the target function \(\psi_t(\cdot)\).

\begin{assumption}\label{assum:lambdaK} 
\begin{enumerate}[itemsep=0pt,label=(\roman*)]  
\item The pairs $(X_i, Y_i)$ for $i = 1, \ldots, n$ are i.i.d. from a distribution $F_{X,Y}$ on $\Rb^d \times \Rb$.
\item $\underline{\lambda}_K > 0$ for all $K$.
\end{enumerate}
\end{assumption}

The above assumption ensures that $Q$ is invertible and is standard in the literature \citep{newey1997convergence, cattaneo2024rosenbaum}. The expressions for $\beta_{t,K}$ and $\hat{\beta}_{t,K}$ can then be simplified to
\[
\beta_{t,K} = Q^{-1} \Er[p_K(\X) \psi_t(\X)], \quad \hat{\beta}_{t,K} = (P^\top P + n \lambda_n I_K)^{-1} P^\top \Ibbm(Y_{[n]} \ge t).
\]

Our goal is to derive a bound for
\[
\Er\left[ \max_{\alpha \in \Lambda_r} \| D^\alpha \hat{\psi}_{t,K} - D^\alpha \psi_t \|_\infty^2 \right].
\]
Such a bound can be directly used to verify Assumption~\ref{assum:estimator}. To this end, we introduce the following assumption, which imposes regularity conditions on the smoothness of the basis functions $p_K$.

\begin{assumption}\label{assum:rate}
Assume $\lambda_n \asymp n^{-c}$ for some $c > 0$ and $K = K_n \to \infty$ as $n \to \infty$. Moreover, assume $\underline{\lambda}_K>\lambda_n$ for all sufficiently large $n$, $K/n\to 0$, $\zeta_{0,K} = o((n/\log n)^{1/4}(\underline{\lambda}_K - \lambda_n)^{1/2})$, 
and $\underline{\lambda}_K^{-1} \zeta_{0,K}^2 \log K = o(n)$.
\end{assumption}

Assumption \ref{assum:rate} is mild. For instance, $\zeta_{r,K} = O(K^{1+r})$ for power series, and $\zeta_{r,K} = O(K^{1/2 + r})$ for Fourier series, splines, compactly supported wavelets, and piecewise polynomials. Moreover, $\underline{\lambda}_K$ is typically bounded away from zero uniformly in $K$; see \citet{newey1997convergence} and \citet{cattaneo2024rosenbaum}. Importantly, we do not require the ``noise'' term $\Ibbm(Y \geq t) - \psi_t(X)$ to be independent of the covariates $X$.

\begin{theorem}[Uniform approximation rate]\label{thm:uni_approx_rate}
Under Assumptions \ref{assum:lambdaK} and \ref{assum:rate}, we have
\[
\Er\left[ \max_{\alpha \in \Lambda_r} \| D^\alpha \hat{\psi}_{t,K} - D^\alpha \psi_t \|_\infty^2 \right] \lesssim  \kappa_{1,n} + \kappa_{2,n} + \kappa_{3,n} + (\vartheta_{r,K}^t)^2,
\]
where
\[
\kappa_{1,n} := \underline{\lambda}_K^{-1} \zeta_{0,K}^2 K n^{-1}, ~~
\kappa_{2,n} := \underline{\lambda}_K^{-4} \left( \underline{\lambda}_K \zeta_{0,K}^2 \log K \cdot n^{-1} + \lambda_n^2 \right), ~~\text{and}~~
\kappa_{3,n} := \zeta_{0,K}^2 \zeta_{r,K}^2 n^{-1}.
\]
\end{theorem}

\section{Simulation studies}

In this section, we present a series of illustrative simulations to demonstrate the performance of the bias-corrected estimator and compare it with the original NN graph-based statistic proposed by \citet{Azadkia21simple}. To this end, we consider the following simulation setup:
\begin{align*}
    \widetilde{X}  = \big(\widetilde{X}^{(1)}, \ldots, \widetilde{X}^{(d)}\big)^\top\sim N(0, I_d), \qquad Z\sim N(0, 1), \qquad \widetilde{Y} = \rho \tilde X^{(1)} + \sqrt{1 - \rho^2} Z.
\end{align*}
In each simulation round, we generate independent copies \( \{(\widetilde{X}_i, \widetilde{Y}_i)\}_{i=1}^n \) from the above multivariate normal distribution. Then we define $X= \big(\Phi(\widetilde{X}^{(1)}), \ldots, \Phi(\widetilde{X}^{(d)})\big)^\top$ and $Y=\Phi(\widetilde{Y})$, where $\Phi$ is the standard normal CDF. 
In this case, by \citet[Propositions~2.4 and 2.7]{ansari2025directextensionazadkia},  we have the closed-form formula:
\[
T(Y,X)\;=\;\frac{3}{\pi}\,\arcsin\!\Big(\frac{1+\rho^{2}}{2}\Big)\;-\;\frac{1}{2}.
\]

We consider \( \rho = 0, 0.3, 0.5, 0.7, 0.9 \), dimensions \( d = 2, 4, 6, 8, 10 \), and sample sizes \( n = 300, 600, 900 \). For the bias correction, we use power series basis functions of polynomial degree equal to $2$. We have set the ridge penalty parameter $\lambda_n = n^{-0.85}$. Variance estimation for both \( \hat{T}_n \) and the bias-corrected estimator \( \hat{T}_n^{\text{bc}} \) is performed via the \( m \)-out-of-\( n \) bootstrap with \( m = \lfloor n^{1/2} \rfloor \), as recommended in \citet[Section 3]{dette2025simple}.

Tables~\ref{tab:sim_d2}--\ref{tab:sim_d10} report the root mean square errors (RMSE) and empirical coverage probabilities (ECP) for both estimators over 1,000 independent repetitions. We consider the nominal significance level \( \alpha = 0.05 \) for the empirical coverage.  The empirical results show that as the dimension or correlation increases, the performance of \( \hat{T}_n^{\text{bc}} \) clearly dominates that of \( \hat{T}_n \). For small dimensions (especially when $d=2$) or correlation (e.g., when $\rho=0$ when no bias is present), the two (original and bias-corrected) correlation estimators have more comparable performance. All these simulations give necessary supplement to the derived theory.

All reproducible code is available in the associated GitHub repository: \url{https://github.com/chenleihaomars/Simulation-for-ACbc}.

{
\renewcommand{\tabcolsep}{20pt}
\renewcommand{\arraystretch}{1}
\begin{table}[H]
\centering
\caption{Root mean squared error (RMSE) and empirical coverage probabilities (ECP) ($d=2$)}
\begin{tabular}{lccccc}
\toprule
\multirow{2}{*}{$\rho$} & \multirow{2}{*}{$n$}
& \multicolumn{2}{c}{RMSE} & \multicolumn{2}{c}{ECP $\alpha=0.05$} \\
\cmidrule(lr){3-4} \cmidrule(lr){5-6} 
& & $\hat{T}_n$ & $\hat{T}_n^{\text{bc}}$ & $\hat{T}_n$ & $\hat{T}_n^{\text{bc}}$  \\
\midrule
  0   & 300 & 0.0628 & 0.0628 & 0.94 & 0.94 \\ 
      & 600 & 0.0441 & 0.0441 & 0.95 & 0.95 \\ 
      & 900 & 0.0350 & 0.0350 & 0.95 & 0.95 \\ 
  0.3 & 300 & 0.0621 & 0.0622 & 0.94 & 0.95 \\ 
      & 600 & 0.0452 & 0.0452 & 0.95 & 0.95 \\ 
      & 900 & 0.0365 & 0.0365 & 0.95 & 0.95 \\ 
  0.5 & 300 & 0.0602 & 0.0602 & 0.95 & 0.95 \\ 
      & 600 & 0.0414 & 0.0414 & 0.95 & 0.95 \\ 
      & 900 & 0.0349 & 0.0349 & 0.95 & 0.95 \\ 
  0.7 & 300 & 0.0553 & 0.0552 & 0.95 & 0.95 \\ 
      & 600 & 0.0392 & 0.0391 & 0.95 & 0.95 \\ 
      & 900 & 0.0323 & 0.0321 & 0.95 & 0.95 \\ 
  0.9 & 300 & 0.0390 & 0.0377 & 0.97 & 0.98 \\ 
      & 600 & 0.0256 & 0.0251 & 0.96 & 0.97 \\ 
      & 900 & 0.0212 & 0.0209 & 0.95 & 0.95 \\ 
\bottomrule
\end{tabular}
\label{tab:sim_d2}
\end{table}
}

{
\renewcommand{\tabcolsep}{20pt}
\renewcommand{\arraystretch}{1}
\begin{table}[H]
\centering
\caption{Root mean squared error (RMSE) and empirical coverage probabilities (ECP) ($d=4$)}
\begin{tabular}{lccccc}
\toprule
\multirow{2}{*}{$\rho$} & \multirow{2}{*}{$n$}
& \multicolumn{2}{c}{RMSE} & \multicolumn{2}{c}{ECP $\alpha=0.05$} \\
\cmidrule(lr){3-4} \cmidrule(lr){5-6} 
& & $\hat{T}_n$ & $\hat{T}_n^{\text{bc}}$ & $\hat{T}_n$ & $\hat{T}_n^{\text{bc}}$  \\
\midrule
  0   & 300 & 0.0628 & 0.0628 & 0.95 & 0.95 \\ 
      & 600 & 0.0448 & 0.0449 & 0.95 & 0.95 \\ 
      & 900 & 0.0367 & 0.0367 & 0.95 & 0.96 \\ 
  0.3 & 300 & 0.0616 & 0.0627 & 0.95 & 0.95 \\ 
      & 600 & 0.0450 & 0.0449 & 0.95 & 0.96 \\ 
      & 900 & 0.0383 & 0.0382 & 0.93 & 0.94 \\ 
  0.5 & 300 & 0.0631 & 0.0629 & 0.94 & 0.94 \\ 
      & 600 & 0.0454 & 0.0449 & 0.94 & 0.94 \\ 
      & 900 & 0.0363 & 0.0357 & 0.95 & 0.96 \\ 
  0.7 & 300 & 0.0663 & 0.0604 & 0.91 & 0.93 \\ 
      & 600 & 0.0457 & 0.0423 & 0.92 & 0.95 \\ 
      & 900 & 0.0366 & 0.0326 & 0.93 & 0.95 \\ 
  0.9 & 300 & 0.0661 & 0.0420 & 0.87 & 0.98 \\ 
      & 600 & 0.0478 & 0.0300 & 0.83 & 0.98 \\ 
      & 900 & 0.0389 & 0.0236 & 0.82 & 0.97 \\ 
\bottomrule
\end{tabular}
\label{tab:sim_d4}
\end{table}
}

{
\renewcommand{\tabcolsep}{20pt}
\renewcommand{\arraystretch}{1}
\begin{table}[H]
\centering
\caption{Root mean squared error (RMSE) and empirical coverage probabilities (ECP) ($d=6$)}
\begin{tabular}{lccccc}
\toprule
\multirow{2}{*}{$\rho$} & \multirow{2}{*}{$n$}
& \multicolumn{2}{c}{RMSE} & \multicolumn{2}{c}{ECP $\alpha=0.05$} \\
\cmidrule(lr){3-4} \cmidrule(lr){5-6} 
& & $\hat{T}_n$ & $\hat{T}_n^{\text{bc}}$ & $\hat{T}_n$ & $\hat{T}_n^{\text{bc}}$  \\
\midrule
  0 & 300 & 0.0627 & 0.0635 & 0.97 & 0.97 \\ 
    & 600 & 0.0484 & 0.0481 & 0.94 & 0.95 \\ 
    & 900 & 0.0383 & 0.0379 & 0.95 & 0.95 \\ 
  0.3 & 300 & 0.0676 & 0.0681 & 0.94 & 0.94 \\ 
    & 600 & 0.0466 & 0.0469 & 0.96 & 0.95 \\ 
    & 900 & 0.0390 & 0.0391 & 0.94 & 0.94 \\ 
  0.5 & 300 & 0.0701 & 0.0678 & 0.93 & 0.94 \\ 
    & 600 & 0.0515 & 0.0466 & 0.92 & 0.95 \\ 
    & 900 & 0.0404 & 0.0361 & 0.93 & 0.95 \\ 
  0.7 & 300 & 0.0826 & 0.0619 & 0.85 & 0.95 \\ 
    & 600 & 0.0598 & 0.0428 & 0.85 & 0.96 \\ 
    & 900 & 0.0502 & 0.0354 & 0.83 & 0.95 \\ 
  0.9 & 300 & 0.1128 & 0.0532 & 0.57 & 0.96 \\ 
    & 600 & 0.0892 & 0.0353 & 0.40 & 0.97 \\ 
    & 900 & 0.0775 & 0.0285 & 0.28 & 0.96 \\ 
\bottomrule
\end{tabular}
\label{tab:sim_d6}
\end{table}
}

{
\renewcommand{\tabcolsep}{20pt}
\renewcommand{\arraystretch}{1}
\begin{table}[H]
\centering
\caption{Root mean squared error (RMSE) and empirical coverage probabilities (ECP) ($d=8$)}
\begin{tabular}{lccccc}
\toprule
\multirow{2}{*}{$\rho$} & \multirow{2}{*}{$n$}
& \multicolumn{2}{c}{RMSE} & \multicolumn{2}{c}{ECP $\alpha=0.05$} \\
\cmidrule(lr){3-4} \cmidrule(lr){5-6} 
& & $\hat{T}_n$ & $\hat{T}_n^{\text{bc}}$ & $\hat{T}_n$ & $\hat{T}_n^{\text{bc}}$  \\
\midrule
  0 & 300 & 0.0654 & 0.0667 & 0.96 & 0.95 \\ 
    & 600 & 0.0475 & 0.0481 & 0.95 & 0.95 \\ 
    & 900 & 0.0391 & 0.0392 & 0.95 & 0.95 \\ 
  0.3 & 300 & 0.0718 & 0.0712 & 0.92 & 0.93 \\ 
    & 600 & 0.0487 & 0.0470 & 0.94 & 0.95 \\ 
    & 900 & 0.0397 & 0.0391 & 0.94 & 0.95 \\ 
  0.5 & 300 & 0.0770 & 0.0689 & 0.89 & 0.94 \\ 
    & 600 & 0.0564 & 0.0467 & 0.89 & 0.96 \\ 
    & 900 & 0.0484 & 0.0391 & 0.89 & 0.96 \\ 
  0.7 & 300 & 0.0974 & 0.0662 & 0.77 & 0.93 \\ 
    & 600 & 0.0771 & 0.0450 & 0.75 & 0.96 \\ 
    & 900 & 0.0663 & 0.0351 & 0.66 & 0.95 \\ 
  0.9 & 300 & 0.1559 & 0.0579 & 0.20 & 0.94 \\ 
    & 600 & 0.1300 & 0.0413 & 0.04 & 0.93 \\ 
    & 900 & 0.1155 & 0.0319 & 0.02 & 0.95 \\ 
\bottomrule
\end{tabular}
\label{tab:sim_d8}
\end{table}
}

{
\renewcommand{\tabcolsep}{20pt}
\renewcommand{\arraystretch}{1}
\begin{table}[H]
\centering
\caption{Root mean squared error (RMSE) and empirical coverage probabilities (ECP) ($d=10$)}
\begin{tabular}{lccccc}
\toprule
\multirow{2}{*}{$\rho$} & \multirow{2}{*}{$n$}
& \multicolumn{2}{c}{RMSE} & \multicolumn{2}{c}{ECP $\alpha=0.05$} \\
\cmidrule(lr){3-4} \cmidrule(lr){5-6} 
& & $\hat{T}_n$ & $\hat{T}_n^{\text{bc}}$ & $\hat{T}_n$ & $\hat{T}_n^{\text{bc}}$  \\
\midrule
  0 & 300 & 0.0693 & 0.0732 & 0.94 & 0.93 \\ 
    & 600 & 0.0480 & 0.0513 & 0.96 & 0.95 \\ 
    & 900 & 0.0406 & 0.0418 & 0.94 & 0.93 \\ 
  0.3 & 300 & 0.0707 & 0.0725 & 0.94 & 0.93 \\ 
    & 600 & 0.0515 & 0.0522 & 0.93 & 0.94 \\ 
    & 900 & 0.0410 & 0.0410 & 0.95 & 0.94 \\ 
  0.5 & 300 & 0.0814 & 0.0782 & 0.90 & 0.91 \\ 
    & 600 & 0.0618 & 0.0526 & 0.86 & 0.92 \\ 
    & 900 & 0.0531 & 0.0410 & 0.87 & 0.95 \\ 
  0.7 & 300 & 0.1085 & 0.0689 & 0.77 & 0.93 \\ 
    & 600 & 0.0896 & 0.0475 & 0.63 & 0.95 \\ 
    & 900 & 0.0802 & 0.0395 & 0.53 & 0.94 \\ 
  0.9 & 300 & 0.1900 & 0.0602 & 0.09 & 0.95 \\ 
    & 600 & 0.1636 & 0.0434 & 0.01 & 0.94 \\ 
    & 900 & 0.1510 & 0.0351 & 0.00 & 0.94 \\ 
\bottomrule
\end{tabular}
\label{tab:sim_d10}
\end{table}
}

\section{Proofs}\label{sec:proofs}

\subsection{Proof for \cref{sec:bias_analysis}}

We introduce some notation. Let $\Sb^{d-1}$ denote the $(d-1)$-dimensional sphere in $\Rb^{d}$ and $B(x,r)\subseteq \Rb^d$ denote the $d$-dimensional open ball in $\Rb^d$ with center $x$ and radius $r$. Let $v_d$ denote the volume of unit Euclidean ball in $\Rb^d$. Denote by $\sigma$ the surface measure on $\Sb^{d-1}$. Let the real number $\Df$ denote the diameter of $\Xc$, i.e., $\Df=\sup_{x,y\in \Xc}\|x-y\|$.

\begin{proof}[Proof of \cref{thm:bias_rate}]

By \cref{lem:bias}, it suffices to show $L=O(n^{-\frac{2}{d}})$. By \cref{ass:G_function}, we have
    \[
        G_{x^\prime}(u)-G_{x}(u)=\nabla_x G_x(u)\cdot (x^\prime-x)+R_x(u,\|x-x^\prime\|),
    \]
    where the remainder $R_x$ satisfies $\sup_{s}|R_x(s,\|x-x^\prime\|)|\le C\|x-x^\prime\|^2$.
    Therefore,
    \[
        \begin{aligned}
            \big|L\big|&=\Big|\Er\big[(X_{N(1)}-X_1)\cdot \Af(X_1)\big]+\Er\int R_{X_1}(u,\|X_{N(1)}-X_1\|)G_{X_1}(u) \dr \mu(u)  \Big|\\
            &\le C\sup_{x^\prime}\|\Af(x^\prime)\|\,\Er\big[\|\Er [X_1-X_{N(1)}\mid X_1]\|\big]+C\Er \|X_1-X_{N(1)}\|^2,
        \end{aligned}
    \]
    where $\Af(X_1)=\int G_{X_1}(t)\nabla_x G_{X_1}(t)\dr \mu(t)$. \cref{ass:G_function} implies that $\sup_{x^\prime}\|\Af(x^\prime)\|<\infty$.
    
    We now analyze the term $\Er\big[\|\Er [X_1-X_{N(1)}\mid X_1]\|\big]$. Similar to \citet{viel2025convergenceratenearestneighbour}, consider the polar representation $X_1-X_{N(1)}=R\Xi$, where $R$ takes values in $(0,\infty)$ and $\Xi$ takes values in the $(d-1)$-sphere $\Sb^{d-1}$. Denote by $\pi_{r,x}(\xi)$ the conditional density of $\Xi$ given $R=r$ and $X_1=x$. Then by \cref{lem:cond_density}, we have
    \[
        \pi_{r,x}(\xi)=\frac{f_X(x+r\xi)\Ibbm(\xi\in A_x(r))}{\int_{A_x(r)} f(x+r\zeta)\dr \sigma(\zeta)},
    \]
    where $\sigma$ is the surface measure on $\Sb^{d-1}$ and $A_x(r)= \{\xi \in \Sb^{d-1}:x+r\xi \in \Xc\}$. Therefore,
    \[
        \Er[\Xi\mid R=r,X_1=x]=\frac{\int_{A_x(r)}\xi\big(f_X(x+r\xi)-f_X(x)\big)\dr \sigma(\xi) +f_X(x)\int_{A_x(r)}\xi \dr \sigma(\xi)}{\int_{A_x(r)}f_X(x+r\xi)\dr \sigma(\xi)}.
    \]
    If $r\le \delta(x)$, then $A_x(r)=\Sb^{d-1}$ and $\int_{A_x(r)}\xi \dr \sigma(\xi)=0$. Since $f_X$ is $\Lc$-Lipschitz and $f_X\ge \pf$, for $r\le \delta(x)$,
    \[
        \big\|\Er[\Xi\mid R=r,X_1=x]\big\|\le \frac{\Lc r\int_{\Sb^{d-1}}\|\xi\|\dr \sigma(\xi)}{\pf\sigma(\Sb^{d-1})}\le Cr.
    \]
    For any $r$ and $x$, we have $\|\Er[\Xi \mid R=r, X=x]\|\le 1$. So for all $x$ and $r$
    \[
        \big\|\Er[\Xi \mid R=r, X=x]\big\|\le Cr\Ibbm(r\leq \delta(x))+\Ibbm(r>\delta(x)).
    \]
    Thus,
    \[
        \begin{aligned}
            \big\|\Er[X_1-X_{N(1)}\mid X_1=x]\big\|&\le \Er\big[R\,\big\|\Er[\Xi\mid R,X_1=x]\big\|\mid X_1=x\big]\\
            &\le C\,\Er[R^2\Ibbm(R\le \delta(x))+R\Ibbm(R>\delta(x))\mid X_1=x].
        \end{aligned}
    \]
    Hence,
    \[
       \Er\big[\|\Er [X_1-X_{N(1)}\mid X_1]\|\big]\lesssim \Er[R^2]+\Er[R\Ibbm(R>\delta(X_1))].
    \]
    In the following, we show $\Er[R\Ibbm(R>\delta(X_1))]=O(n^{-\frac{2}{d}})$. First note that  
    \[
        \Er[R\Ibbm(R>\delta(X_1))]=\Er(R-\delta(X_1))^+ +\Er\delta(X_1)\Ibbm(\delta(X_1)<R)=: \Ec_1+\Ec_2, 
    \]
    where $(R-\delta(X_1))^+=\max\{R-\delta(X_1),0\}$. We analyze $\Ec_1$ and $\Ec_2$ individually. First, note that
    \[
        \Ec_1=\int_0^\infty \Prb(\delta(X_1)< r<R)\dr r, \text{ and } \Prb(\delta(X_1)< r <R)=\Er[\Ibbm(\delta(X_1)< r)\Prb(R>r\mid X_1)].
    \]
    When $r\le r_0$ for a sufficiently small $r_0>0$, since \cref{lem:Lip_boundary_interior_nondegeneracy} establishes for all $x\in \Xc$
    \[
        p_x(r)\coloneqq \int_{B(x,r)\cap \Xc} f\ge \pf |B(x,r)\cap \Xc| \ge \pf c^d_0v_dr^d,
    \]
    it holds for all $x\in \Xc$ that
    \[
        \Prb(R>r\mid X_1=x)=(1-p_x(r))^{n-1}\le \exp(-Cnr^d).
    \]
    So 
    \[
        \Ec_1\le \int_0^{r_0} \Prb(\delta(X_1)< r)\exp(-Cnr^d)\dr r+ \int_{r_0}^{\Df} \Prb(R>r,\delta(X_1)< r)\dr r =: \Ec_{1,1}+\Ec_{1,2},
    \]
    where $\Df$ is the diameter of $\Xc$.
    For $\Ec_{1,1}$, we have by \cref{lem:Lip_boundary_linear_bound,lem:integral},
    \[
        \Ec_{1,1}\le C\int_0^{r_0}r\exp(-Cnr^d)\dr r\lesssim n^{-2/d}.
    \]
    For $\Ec_{1,2}$, since $\Prb(R>r,\delta< r)\le \Prb(R>r_0)\le \exp(-Cnr_0^d)$, we have
    \[
        \Ec_{1,2}\le \int_{r_0}^\Df \exp(-Cnr_0^d)\dr r\le (\Df-r_0)\exp(-Cnr_0^d)=o(n^{-2/d}).
    \]
    This shows $\Ec_1=O(n^{-2/d})$. Now we show $\Ec_2=O(n^{-2/d})$. Note that 
    \[
    \begin{aligned}
        \Ec_{2}&\le \|f_X\|_{\infty}\int_{\Xc}\delta(x)\Prb(R>\delta(x)\mid X_1=x)\dr x \\
        &\le C\|f_X\|_{\infty}\int_{\{0\le \delta(x)\le r_0\}}\delta(x)\exp(-Cn\delta(x)^d)\dr x+C\|f_X\|_{\infty}\exp(-Cnr_0^d)\int_{\{r_0< \delta(x)\le \Df\}} \delta(x)\dr x\\
        &\lesssim  \int_{\{0\le \delta(x)\le r_0\}}\delta(x)\exp(-Cn\delta(x)^d)\dr x+ \exp(-Cnr_0^d)\Df|\Xc|=\widetilde{\Ec}_{2}+o(n^{-2/d}),
    \end{aligned}
    \]
    where $\widetilde{\Ec}_{2}=\int_{\{0\le \delta(x)\le r_0\}}\delta(x)\exp(-Cn\delta(x)^d)\dr x$. Set 
    \[
    h(t)\coloneqq te^{-Cnt^d} 
    \]
    and monotone increasing function 
    \[
    m(t)\coloneqq |\{x\in \Xc:0\le \delta(x)\le t\}| ~~\text{ for }t\in [0,r_0]. 
    \]
    Then integration by parts yields 
    \[  
            \widetilde{\Ec}_2=\int_0^{r_0} h(t)\dr m(t)=h(r_0)m(r_0)-h(0)m(0)-\int_0^{r_0}m(t)h^\prime(t)\dr t.
    \]
    Note that $h(r_0)m(r_0)\le Cr_0^2e^{-Cnr_0^d}=o(n^{-2/d})$, $h(0)m(0)=0$ and 
    \[
        \int_0^{r_0}m(t)h^\prime(t)\dr t\le C\int_0^{r_0} e^{-Cnt^d}(t+t^{d+1}n)\dr t=O(n^{-2/d})
    \]
    by \cref{lem:Lip_boundary_linear_bound,lem:integral}. Hence, $\widetilde{\Ec}_{2}=O(n^{-2/d})$.
\end{proof}

\begin{proof}[Proof of \cref{thm:bias_expansion}]
    By \cref{lem:bias}, we can write
    \[
        \Er[\widehat{T}_n]-T=6L-\frac{6}{n+1}E_{1,n}+\frac{6n}{n^2-1}E_{2,n}-\frac{3}{n-1},
    \]
    where $E_{1,n}=\Er[\Ibbm(Y_2\le Y_1\wedge Y_{N(1)})]$ and $E_{2,n}=\Er[\Ibbm(Y_1\le Y_1\wedge Y_{N(1)})]$. Note that, as $n\to \infty$,
    \[
        -\frac{6}{n+1}E_{1,n}+\frac{6n}{n^2-1}E_{2,n}-\frac{3}{n-1}=n^{-1}(-6E_{1,n}+6E_{2,n}-3)+o(n^{-1})
    \]
    and we have the following Lemma:
    \begin{lemma}\label{lem:E}
       It holds true that $E_{1,n}\to \int \Er[G_X(t)^2]\dr \mu(t)$ and $E_{2,n}\to 1/2$ as $n\to \infty$.
    \end{lemma}

    It remains to analyze the first term $L$. For that purpose, fix a positive sequence 
    \[
    r_n=n^{-a} \text{ with }a=\frac{4d+5}{4d^2+6d}.
    \]
    Write 
    \[
    \Xc_r\coloneqq \{x\in \Xc:\delta(x)\ge r\} ~~\text{for any } r>0 
    \]
    and
    \[
        \begin{aligned}
            L=&\, \Er \Big[\int \big(G_{X_{N(1)}}(t)-G_{X_1}(t)\big)G_{X_1}(t)\dr \mu(t)\Big]\\
            =&\, \Er \Big[\int \big(G_{X_{N(1)}}(t)-G_{X_1}(t)\big)G_{X_1}(t)\dr \mu(t) \cdot \Ibbm(X_1\in \Xc_{r_n})\Big]\\
            & +\Er \Big[\int \big(G_{X_{N(1)}}(t)-G_{X_1}(t)\big)G_{X_1}(t)\dr \mu(t) \cdot \Ibbm(X_1\in \Xc\setminus \Xc_{r_n})\Big]\\
            =: &\, L_n^{\mathrm{int}}+L_n^{\mathrm{bd}}.
        \end{aligned}
    \]
    The goal is to show 
    \[
    L_n^{\mathrm{int}}=\widetilde{\Cf}_1n^{-\frac{2}{d}}+o(n^{-\frac{2}{d}}) ~~\text{ and }~~L_n^{\mathrm{bd}}=\widetilde{\Cf}_3 n^{-\frac{2}{d}}+o(n^{-\frac{2}{d}}),
    \]
    where $\widetilde{\Cf}_1=-\frac{1}{6}\Cf_1$ and $\widetilde{\Cf}_3=-\frac{1}{6}\Cf_3$.

    \Step[step1:int]{Analyze $L_n^{\mathrm{int}}$} By \cref{assum:ass_5}, Taylor expansion gives, for small $0<r< r_n$,
    \begin{equation}\label{eq:G}
        G_{x+r\xi}(t)-G_x(t)=r\nabla_x G_x(t)\cdot \xi +\frac{1}{2}r^2\xi^\top\nabla_x^2 G_x(t) \xi +O(r^{2+\beta})
    \end{equation}
    uniformly in $x\in \Xc_{r_n}, t\in \supp(\mu), \xi \in \Sb^{d-1}$, i.e.,
    \[
        \sup_{x\in \Xc_{r_n},t\in \supp(\mu),\xi\in \Sb^{d-1}}\big|G_{x+r\xi}(t)-G_x(t)-\big(r\nabla_x G_x(t)\cdot \xi +\frac{1}{2}r^2\xi^\top\nabla_x^2 G_x(t)\xi\big)\big|\le Cr^{2+\beta}.
    \]
    By \cref{assum:ass_5}, the Taylor expansion of $f_X$ at $x\in \Xc_{r_n}$ as $r\downarrow 0$ gives:
    \[
        f_X(x+r\xi)=f_X(x)+r\nabla f_X(x)\cdot \xi+R_1(x,r,\xi),
    \]
    and since $\int_{\Sb^{d-1}}\xi \dr \sigma(\xi)=0$,
    \[
    \begin{aligned}
        \int_{\Sb^{d-1}} f_X(x+r\xi)\dr\sigma(\xi)&=f_X(x)\sigma(\Sb^{d-1})+r\nabla f_X(x)\cdot \int_{\Sb^{d-1}}\xi \dr \sigma(\xi)+R_2(x,r)\\
        &=f_X(x)\sigma(\Sb^{d-1})+R_2(x,r),
    \end{aligned}
    \]
    where $|R_1(x,r,\xi)|\le Cr^{1+\alpha}$ and $|R_2(x,r)|=|\int_{\Sb^{d-1}}R_1(x,r,\xi)\dr \sigma(\xi)|\le C\sigma(\Sb^{d-1})r^{1+\alpha}$. Therefore, by \cref{lem:cond_density},
    \[
        \pi_{r,x}(\xi)=\frac{f_X(x)+r\nabla f_X(x)\cdot \xi +R_1(x,r,\xi)}{f_X(x)\sigma(\Sb^{d-1})+R_2(x,r)}.
    \]
    Note that $\int_{\Sb^{d-1}}\xi R_1(x,r,\xi)\dr \sigma(\xi)=O(r^{1+\alpha})$ and $\int_{\Sb^{d-1}} f_X(x) \xi \dr \sigma(\xi)=0$. Write 
    \[
    \kappa(x,r)\coloneqq \frac{R_2(x,r)}{\sigma(\Sb^{d-1})f_X(x)}. 
    \]
    Then it yields
    \[
        \begin{aligned}
            \Er[\Xi\mid R=r,X_1=x]&=\int_{\Sb^{d-1}}\xi \pi_{r,x}(\xi)\dr \sigma(\xi)\\
            &=\frac{r \int_{\Sb^{d-1}} \xi(\nabla f_X(x)\cdot \xi)\dr \sigma(\xi)+O(r^{1+\alpha})}{f_X(x)\sigma(\Sb^{d-1})+R_2(x,r)}\\
            &=\frac{d^{-1}r\nabla \log f_X(x)+O(r^{1+\alpha})}{1+\kappa(x,r)},
        \end{aligned}
    \]
    since $\int_{\Sb^{d-1}} \xi(\nabla f_X(x)\cdot \xi) \dr\sigma(\xi)=\int_{\Sb^{d-1}} \xi\xi^\top \dr\sigma(\xi) \nabla f_X(x)=\frac{\sigma(\Sb^{d-1})}{d} \nabla f_X(x)$. Since $\kappa(x,r)=O(r^{1+\alpha})$ uniformly in $x\in \Xc_{r_n}$, we have
    \begin{equation}\label{eq:xi}
        \begin{aligned}
            \Er[\Xi\mid R=r,X_1=x]&=d^{-1}r\nabla\log f_X(x)\big(1-\kappa(x,r)+O(\kappa(x,r)^2)\big)\\
            &=d^{-1}r\nabla\log f_X(x)+O(r^{1+\alpha}).
        \end{aligned}
    \end{equation}
    Similarly, we have 
    \[
        \Er[\Xi\Xi^\top\mid R=r,X_1=x]=\frac{d^{-1}I_d}{1+\kappa(x,r)}+\frac{R_3(x,r)}{\sigma(\Sb^{d-1})f_X(x)(1+\kappa(x,r))},
    \]
    where 
    \[
    R_{3}(x,r):=\int_{\Sb^{d-1}}\xi \xi^\top R_1(x,r,\xi)\dr \sigma(\xi). 
    \]
    Note that $\|R_3(x,r)\|_2\le C\sigma(\Sb^{d-1})r^{1+\alpha}$. So, uniformly in $x\in \Xc_{r_n}$
    \begin{equation}\label{eq:xi2}
        \Er[\Xi\Xi^\top\mid R=r,X_1=x]=d^{-1}I_d+O(r^{1+\alpha}).
    \end{equation}
    Combining \cref{eq:G,eq:xi,eq:xi2} gives: uniformly in $x\in \Xc_{r_n}$
    \[
        \begin{aligned}
            &\Er\Big[\int\big(G_{X_{N(1)}}(t)-G_{X_1}(t)\big)G_{X_1}(t)\dr \mu(t) \Ibbm(R\le r_n)\mid X_1=x\Big]\\
            =&d^{-1}\Er[R^2\Ibbm(R\le r_n)\mid X_1=x]\big(\nabla \log f_X(x)\cdot \Af(x)+\frac{1}{2}\Tr(\Bf(x))\big)+o(n^{-\frac{2}{d}}).
        \end{aligned}
    \]
    We have
    \[
        \Er[R^2\Ibbm(R\le r_n)\mid X_1=x]=\Er[R^2\mid X_1=x]+o(n^{-\frac{2}{d}})
    \]
    uniformly in $x\in \Xc_{r_n}$, since for every $x\in \Xc_{r_n}$
    \[
        \begin{aligned}
            \Er[R^2\Ibbm(R>r_n)\mid X_1=x]&=r_n^2\Prb(R>r_n\mid X_1=x) +2\int_{r_n}^\Df r\Prb(R>r\mid X_1=x)\dr r\\
            &\lesssim r_n^2\exp(-Cnr^d_n)+(\Df^2-r_n^2)\exp(-Cnr^d_n)=o(n^{-\frac{2}{d}}),
        \end{aligned} 
    \]
    where we use the fact that $r_n\downarrow 0$ with $r_n^{-d}=o\big(\frac{n}{\log n^{2/d}}\big)$.
    By the boundedness of function $G$, we have
    \[
        \begin{aligned}
            &\Er\Big[\int\big(G_{X_{N(1)}}(t)-G_{X_1}(t)\big)G_{X_1}(t)\dr \mu(t) \Ibbm(R> r_n)\mid X_1=x\Big]\\
            \lesssim& \Prb(R>r_n\mid X_1=x)\le \exp(-Cnr_n^{d}) =o(n^{-\frac{2}{d}}).
        \end{aligned}
    \]
    This implies
    \[
        \begin{aligned}
            &\Er\Big[\int\big(G_{X_{N(1)}}(t)-G_{X_1}(t)\big)G_{X_1}(t)\dr \mu(t) \Ibbm(R\le r_n)\mid X_1=x\Big]\\
        =&\Er\Big[\int\big(G_{X_{N(1)}}(t)-G_{X_1}(t)\big)G_{X_1}(t)\dr \mu(t) \mid X_1=x\Big]+o(n^{-\frac{2}{d}})
        \end{aligned}
    \]
    uniformly in $x\in \Xc_{r_n}$, and therefore
    \[
        L_{n}^{\mathrm{int}}=\int_{\Xc_{r_n}}f_X(x)d^{-1}\big(\nabla \log f_X(x)\cdot \Af(x)+\frac{1}{2}\Tr(\Bf(x)) \big) \Er[R^2\mid X_1=x]\dr x+o(n^{-\frac{2}{d}}).
    \]
    Note that the following holds:
    \begin{lemma}\label{lem:Er^2|x}
        We have $\Er[R^2\mid X_1=x]=\Gamma(1+2/d)(v_df_X(x)n)^{-\frac{2}{d}}+o(n^{-\frac{2}{d}})$ uniformly in $x\in \Xc_{r_n}$ provided $r_n\downarrow 0$ with $n r_n^d\to \infty$.
    \end{lemma} 
    Plugging it in gives
    \[
        L_n^{\mathrm{int}}=\Gamma(1+2/d)(v_dn)^{-\frac{2}{d}}\int_{\Xc_{r_n}} I(x)\dr x +o(n^{-\frac{2}{d}}),
    \]
    where 
    \[
    I(x):=f_X(x)^{1-\frac{2}{d}}d^{-1}\big(\nabla \log f_X(x)\cdot \Af(x)+\frac{1}{2}\Tr(\Bf(x)) \big). 
    \]
    The integrand $I(x)$ is continuous and bounded on $\Xc$, so $\int_{\Xc_{r_n}} I(x)\dr x\to \int_{\Xc} I(x)\dr x<\infty$ as $n\to \infty$. Hence,
    \[
        L_n^{\mathrm{int}}=\widetilde{\Cf}_1n^{-\frac{2}{d}}+o(n^{-\frac{2}{d}}).
    \]

    \Step[step2:bd]{Analyze $L_n^{\mathrm{bd}}$} 

    Set $I(X_{[n]})\coloneqq \int\big(G_{X_{N(1)}}(t)-G_{X_{1}}(t)\big)G_{X_{1}}(t)\dr \mu(t)$. Then 
    \[
    \begin{aligned}
        L^{\mathrm{bd}}_n&=\int_{\Xc\setminus \Xc_{r_n}}f_X(x)\Er[I(X_{[n]})\Ibbm(R\le \delta(x))\mid X_1=x]\dr x+\int_{\Xc\setminus \Xc_{r_n}}f_X(x)\Er[I(X_{[n]})\Ibbm(R> \delta(x))\mid X_1=x]\dr x\\
   & =: L^{\mathrm{bd}}_{n,1}+L^{\mathrm{bd}}_{n,2}.
    \end{aligned}
    \]
    We show $ L^{\mathrm{bd}}_{n,1}=o(n^{-2/d})$. Similar to the argument before (note that on $\{R\le \delta(x)\}$ the ball $B(x,R)\subseteq \Xc$), we have 
    \[
        \big|\Er[I(X_{[n]})\Ibbm(R\le \delta(x))\mid X_1=x]\big|\lesssim \Er[R^2\mid X_1=x]\lesssim n^{-\frac{2}{d}}
    \]
    uniformly for all $x\in \Xc$. Therefore, by \cref{lem:Lip_boundary_linear_bound},
    \[
        \int_{\Xc\setminus \Xc_{r_n}} f_X(x)\big|\Er[I(X_{[n]})\Ibbm(R\le \delta(x))\mid X_1=x]\big|\dr x \lesssim n^{-\frac{2}{d}}r_n=o(n^{-\frac{2}{d}}).
    \]
    We analyze $ L^{\mathrm{bd}}_{n,2}$. Note that 
    \[
        L^{\mathrm{bd}}_{n,2}=\tilde{L}_{n,2}^{\mathrm{bd}}+\int_{\Xc\setminus \Xc_{r_n}}f_X(x)\Er[I(X_{[n]})\Ibbm(R>r_n)\mid X_1=x]\dr x=\tilde{L}_{n,2}^{\mathrm{bd}}+o(n^{-2/d}),
    \]
    where 
    \[
        \tilde{L}_{n,2}^{\mathrm{bd}}=\int_{\Xc\setminus \Xc_{r_n}}f_X(x)\Er[I(X_{[n]})\Ibbm(\delta(x)<R\le r_n)\mid X_1=x]\dr x.
    \]
    Indeed, since $p_r(x)\ge \pf |\Xc\cap B(x,r)|\ge \pf c_0 v_d (r_n\wedge r_0)^d$ for all $x\in \Xc$ and $r\in [r_n,\infty)$, we have uniformly in $x$
    \[
        \Er[I(X_{[n]})\Ibbm(R>r_n)\mid X_1=x]\lesssim
        (1-p_{r_n}(x))^{n-1}\le \exp(-(n-1)p_{r_n}(x))\le \exp(-c_1 n r_n^d)=o(n^{-2/d}).
    \]
    Now, we analyze $\tilde{L}_{n,2}^{\mathrm{bd}}$. By  \cref{lem:cond_density}, we have
    \[
        \Er[\Xi\mid R=r,X_1=x]=\frac{\int_{A_x(r)}\xi f_X(x+r\xi)\dr \sigma(\xi)}{\int_{A_x(r)} f_X(x+r\xi)\dr \sigma(\xi)}.
    \]
    Define 
    \[
    \Sc_n\coloneqq \Big\{(x,r):x\in \Xc\setminus \Xc_{r_n},0\le \delta(x)\le r\le r_n\Big\}. 
    \]
    Using the Lipschitz continuity and strict positivity of $f_X$ on the compact support and the fact that $\sigma(A_x(r))$ is bounded away from zero uniformly on $\Sc_n$, it follows that
    \[
         \Er[\Xi\mid R=r,X_1=x]=\frac{\int_{A_x(r)}\xi\dr \sigma(\xi)}{\sigma(A_x(r))}+O(r)
    \]
    uniformly for $(x,r)\in \Sc_n$. Define 
    \[
    C_{x}(r)\coloneqq \Big\{\xi\in \Sb^{d-1}:\xi\cdot \nabla\delta(x)\ge -\delta(x)/r\Big\}. 
    \]
    In angular coordinates, the actual feasible direction set $A_x(r)$ differs only slightly in measure from the spherical cap $C_x(r)$, which is justified by the following lemma.
    \begin{lemma}\label{lem:symmetric_difference}
         We have $\sigma(A_x(r)\sd C_{x}(r))=o(1)$ as $n\to \infty$ uniformly for a.e.\ $(x,r)\in \Sc_n$.
    \end{lemma}
    Since the function $\xi\mapsto \xi$ is bounded and both $\sigma(A_x(r))$ and $\sigma(C_x(r))$ are uniformly bounded away from zero, we have uniformly for a.e.\ $(x,r)\in \Sc_n$
     \[
        \Er[\Xi\mid R=r,X_1=x]=\sigma(C_{x}(r))^{-1}\int_{C_{x}(r)}\xi \dr \sigma(\xi)+o(1).
    \]
    Define 
    \[
    \Mf_d(\tau)\coloneqq \sigma(C_x(r))^{-1}\int_{C_x(r)}\xi \cdot \nabla \delta(x)\dr \sigma(\xi)~~ \text{ for }\tau=\delta(x)/r\in[0,1]. 
    \]
    Then we can write 
    \[
    \sigma(C_x(r))^{-1}\int_{C_x(r)}\xi \dr \sigma(\xi)=\Mf_{d}(\delta(x)/r)\nabla \delta(x).
    \] 
    Plugging the expression for $\int_{C_{x}(r)}\xi \dr \sigma(\xi)$ into the expression of $\Er[\Xi\mid R=r,X_1=x]$ gives, uniformly in $x\in \Xc\setminus \Xc_{r_n}$,
    \[
        \begin{aligned}
            &\Er[I(X_{[n]})\Ibbm(\delta(x)<R\le r_n)\mid X_1=x]\\
            =&\Er[R\ \Mf_d(\delta(x)/R)\Gf(x)\Ibbm(\delta(x)<R\le r_n)\mid X_1=x]+\Er[o(R)\Ibbm(\delta(x)<R\le r_n)\mid X_1=x]\\
            =:& \Upsilon_1(x)+\Upsilon_2(x).
        \end{aligned}       
    \]
    Note that 
    \[             
        \begin{aligned}  \tilde{L}^{\mathrm{bd}}_{n,2}&=\int_{\{0\le \delta(x)\le r_n\}}f_X(x)\Er[I(X_{[n]})\Ibbm(\delta(x)<R\le r_n)\mid X_1=x]\dr x\\
        &=\int_{\{0\le \delta(x)\le r_n\}}f_X(x) \Upsilon_1(x) \dr x+\int_{\{0\le \delta(x)\le r_n\}}f_X(x) \Upsilon_2(x)\dr x\\
        &=: \tilde{L}^{\mathrm{bd}}_{n,2,1}+\tilde{L}^{\mathrm{bd}}_{n,2,2}.
        \end{aligned}
    \]
    
    The goal is to show $\tilde{L}^{\mathrm{bd}}_{n,2,2}=o(n^{-2/d})$.
     We have for $x\in \Xc\setminus \Xc_{r_n}$
    \[
        \begin{aligned}
             \Er[R\Ibbm(\delta(x)<R\le r_n)\mid X_1=x]&\le  \delta(x)\Prb(R>\delta(x)\mid X_1=x)+\int_{\delta(x)}^\Df \Prb(R>r\mid X_1=x)\dr r, \\
        &\lesssim \delta(x)\exp(-Cn\delta(x)^d)+n^{-\frac{1}{d}}\exp(-Cn\delta(x)^d).
        \end{aligned} 
    \]
    Note that similar to the analysis of $\widetilde{\Ec}_2$ in the proof of \cref{thm:bias_rate}, we have
    \[
        \begin{aligned}
        \int_{\{0\le \delta(x)\le r_n\}}\delta(x)\exp(-Cn\delta(x)^d)\dr x&=O(n^{-2/d}) \quad \text{ and } \\
         n^{-1/d}\int_{\{0\le \delta(x)\le r_n\}}\exp(-Cn\delta(x)^d)\dr x&=O(n^{-2/d}).
        \end{aligned}
    \]
    Hence, $\tilde{L}^{\mathrm{bd}}_{n,2,2}=o(n^{-2/d})$.

    Now we analyze the main term $\tilde{L}^{\mathrm{bd}}_{n,2,1}$. Note that
    \[
            \tilde{L}^{\mathrm{bd}}_{n,2,1}=\int_{\Xc\setminus \Xc_{r_n}}f_X(x)\Gf(x)\int_{\delta(x)}^{r_n} \int_{\Sb^{d-1}} r\Mf_d(\delta(x)/r)g_x(r,\xi)\dr \sigma(\xi) \dr r\dr x,
    \]
    where $g_x(r,\xi)$ is the conditional density of $(R,\Xi)$ given $X_1=x$ in \cref{lem:cond_density} and 
    \begin{equation}\label{eq:g}
        g_x(r)=\int_{\Sb^{d-1}} g_x(r,\xi)\dr \sigma(\xi)=(n-1)r^{d-1}(1-p_r(x))^{n-2}\int_{A_x(r)}f_X(x+r\xi)\dr \sigma(\xi).
    \end{equation}
    By \cref{lem:symmetric_difference}, we have, uniformly in $x\in \Xc\setminus \Xc_{r_n}$,
    \begin{equation}\label{eq:int_f}
        \int_{A_x(r)}f_X(x+r\xi) \dr \sigma(\xi)=f_X(x)\sigma(C_x(\delta(x)/r))(1+o(1)).
    \end{equation}

    \begin{lemma}\label{lem:ball_asymp}
        It holds true that, as $n\to \infty$,
        \[
        \sup_{(x,r)\in \Sc_n}\Bigg|\frac{|B(x,r)\cap \Xc|-v_d\Ff_d(\delta(x)/r)r^d}{r^d}\Bigg|=o(1),
        \] 
        where $\Ff_d(\tau):=\frac{1}{2}\Big(1+I_{\tau^2}\big(\frac{1}{2},\frac{d+1}{2}\big)\Big)$ and $I_{\cdot}(\cdot,\cdot)$ denotes the regularized incomplete beta function.
    \end{lemma}
    Write $\rho_x(\tau)=nf_X(x)v_d\Ff_d(\tau)$.  Note that $nr_n^{d+1}=o(1)$ and $nr_n^{2d}=o(1)$.
    \cref{lem:ball_asymp} implies that, as $n\to \infty$,
    \begin{equation}\label{eq:p}
        \sup_{(x,r)\in \Sc_n}\big|(1-p_{r}(x))^{n-2}-\exp(-\rho_x(\delta(x)/r)r^d)\big|=o(1).
    \end{equation}
    Plugging \cref{eq:int_f,eq:p} into \cref{eq:g} gives, uniformly in $\Sc_n$,
    \[
        g_x(r)=(n-1)r^{d-1}\exp(-\rho_x(\delta(x)/r)r^d)f_X(x)\sigma(C_x(\delta(x)/r))(1+o(1)).
    \]
    The above derivation yields $\tilde{L}^{\mathrm{bd}}_{n,2,1}=\varpi(1+o(1))$, where
    \[
        \varpi=\int_{\{0\le \delta(x)\le r_n\}}f_X(x)\Gf(x)\int_{\delta(x)}^{r_n}r\Mf_d(\delta(x)/r)(n-1)r^{d-1}\exp(-\rho_x(\delta(x)/r)r^d)f_X(x)\sigma(C_x(\delta(x)/r))\dr r \dr x.
    \]
    The next step is to deduce the asymptotic expression of $\varpi$, which is of the order $n^{-2/d}$. Note that
    \[
        \begin{aligned}
            \varpi&=\int_{\{0\le \delta(x)\le r_n\}}\Gf(x)f_X(x)\int_{\delta(x)/r_n}^{1}\Mf_d(\tau)(n-1)\delta(x)^{d+1}\tau^{-(d+2)}\exp(-\rho_x(\tau)\delta(x)^d\tau^{-d})f_X(x)\sigma(C_x(\tau))\dr \tau\dr x\\
            &=(n-1)\int_0^1\Mf_d(\tau)\sigma(C(\tau))\tau^{-(d+2)}\int_{\{0\le \delta(x)\le r_n\tau\}} \delta(x)^{d+1}\exp(-\rho_x(\tau)\delta(x)^d\tau^{-d})f_X^2(x)\Gf(x)\dr x\dr \tau,
        \end{aligned}
    \]
    where we apply the change of variable $\tau=\delta(x)/r$ in the first equality and use the Fubini theorem in the second equality, and denote $\sigma(C_x(\tau))$ by $\sigma(C(\tau))$ as $\sigma(C_x(\tau))$ does not depend on $x$. We then have the following lemma.

    \begin{lemma}\label{lem:approximate_bounday_proj}
            We have $\varpi=\widetilde{\Cf}_3n^{-2/d}+o(n^{-2/d})$.
    \end{lemma}
    
    This completes the proof.
\end{proof}

\subsection{Proofs for \cref{sec:theory}}

We start by introducing the additional notation used in the following. In the following, let $\|\cdot\|_2$ be the spectral (operator) norm of a matrix, $\Tr(M)$ denote the trace of matrix $M$, and $\Diag(m_1,\ldots,m_n)$ represent an $n$-dimensional diagonal matrix with diagonal elements $m_1,\ldots,m_n$. For symmetric matrices $M_1$ and $M_2$, we write $M_1\preceq M_2$ if $M_2-M_1$ is a positive semi-definite matrix. Let $\|\cdot\|_{L^2}$  represent the $L^2(F_X)$-norm of functions.

\begin{proof}[Proof of \cref{thm:convergence_B}]
Write
\[
  H_t          \coloneqq G_{\X_1}(t)\,G_{\X_{N(1)}}(t) - G_{\X_1}^2(t),
  \quad
  \hat{H}_t    \coloneqq \hat{G}_{\X_1}(t)\,\hat{G}_{\X_{N(1)}}(t)
                   - \hat{G}_{\X_1}^2(t),
\]

Note that
\[
\begin{aligned}
 L-\hat{L}_n
    &= \int \Er\big[H_t\big]\,\dr\mu(t)
      - \int \hat{\Er}\big[\hat{H}_t\big]\,\dr\hat{\mu}(t)\\
    &= \int \Er\big[H_t - \hat{H}_t\big]\,\dr\mu(t)
      + \int \big(\Er\big[\hat{H}_t\big] - \hat{\Er}\big[\hat{H}_t\big]\big)
        \,\dr\mu(t)
      + \bigg(
          \int \hat{\Er}\big[\hat{H}_t\big]\,\dr\mu(t)
          - \int \hat{\Er}\big[\hat{H}_t\big]\,\dr\hat{\mu}(t)
        \bigg)\\
    &=: B_1 + B_2 + B_3.
\end{aligned}
\]
The proof proceeds with the individual analysis of $B_1$, $B_2$, and $B_3$.

\Step[step1:boundingB1]{Bound $B_1$}
    
We have
\[
  \begin{aligned}
    B_1
    &= \int 
       \Er\bigl[\bigl(G_{\X_1}(t)-\hat{G}_{\X_1}(t)\bigr)\,
                \bigl(G_{\X_{N(1)}}(t)-G_{\X_1}(t)\bigr)\bigr]
       \,\dr\mu(t)\\
    &\quad
      + \int 
        \Er\bigl[\hat{G}_{\X_1}(t)\,
                  \bigl(\bigl(G_{\X_{N(1)}}(t)-\hat{G}_{\X_{N(1)}}(t)\bigr)
                      - \bigl(G_{\X_1}(t)-\hat{G}_{\X_1}(t)\bigr)\bigr)\bigr]
        \,\dr\mu(t)\\
    &= \int 
       \Er\bigl[\bigl(G_{\X_1}(t)-\hat{G}_{\X_1}(t)\bigr)\,
                \bigl(G_{\X_{N(1)}}(t)-G_{\X_1}(t)\bigr)\bigr]
       \,\dr\mu(t)\\
    &\quad
      + \int 
        \Er\bigl[G_{\X_1}(t)\,
                  \bigl(\bigl(G_{\X_{N(1)}}(t)-\hat{G}_{\X_{N(1)}}(t)\bigr)
                      - \bigl(G_{\X_1}(t)-\hat{G}_{\X_1}(t)\bigr)\bigr)\bigr]
        \,\dr\mu(t)\\        
     &\quad
      + \int 
        \Er\bigl[(\hat G_{\X_1}(t)-G_{\X_1}(t))\,
                  \bigl(\bigl(G_{\X_{N(1)}}(t)-\hat{G}_{\X_{N(1)}}(t)\bigr)
                      - \bigl(G_{\X_1}(t)-\hat{G}_{\X_1}(t)\bigr)\bigr)\bigr]
        \,\dr\mu(t)\\          
    &=: B_{11} + B_{12}+B_{13}.
  \end{aligned}
\]
The goal is to show $|B_{1j}|=o(n^{-\frac{1}{2}})$ for $j=1,2,3$.

For the first term $B_{11}$, by Assumptions \ref{assum:x}-\ref{assum:estimator}, we have
\[
  \begin{aligned}
    |B_{11}|
    &\le \int 
        \Er\bigl[\,|G_{\X_1}(t)-\hat{G}_{\X_1}(t)|\;
                  |G_{\X_{N(1)}}(t)-G_{\X_1}(t)|\,\bigr]
        \,\dr\mu(t)\\
    &\lesssim \int 
        \Er\bigl[\big\|G_{\cdot}(t)-\hat{G}_{\cdot}(t)\big\|_{\infty}\;
                  \sup_{t}\max_{\alpha\in \Lambda_1}\|D^\alpha G_{\cdot}(t)\|_{\infty}\;
                  \big\|\X_{{N}(1)}-\X_1\big\|\bigr]
        \,\dr\mu(t)\\
    &\lesssim \sup_{t}\max_{\alpha\in \Lambda_1}\|D^\alpha G_{\cdot}(t)\|_{\infty}\,
      \Bigl(\int 
        \Er^{1/2}\bigl[\big\|G_{\cdot}(t)-\hat{G}_{\cdot}(t)\big\|_{\infty}^2\bigr]
        \,\dr\mu(t)\Bigr)\,
      \Er^{1/2}\bigl[\big\|\X_{N(1)}-\X_1\big\|^2\bigr]\\
    &= o\bigl(n^{-\frac12}\bigr),
  \end{aligned}
\]
where we use the Cauchy-Swartz inequality in the third inequality.

Now we consider $B_{12}$. As $\big|G_X(t)\big|$ is bounded, we have
\[
    |B_{12}|\lesssim \int \Er\big[ \big| \bigl(G_{\X_{N(1)}}(t)-\hat{G}_{\X_{N(1)}}(t)\bigr)
                      - \bigl(G_{\X_1}(t)-\hat{G}_{\X_1}(t)\bigr)\big| \big] \dr \mu(t)=: \Xi.
\]
It suffices to show $\Xi=o(n^{-\frac{1}{2}})$. By the Taylor expansions of $G_{\X_{N(1)}}(t)-G_{\X_1}(t)$ and $\hat{G}_{\X_{N(1)}}(t)-\hat{G}_{\X_1}(t)$, we have, with $k=\lfloor d/2 \rfloor +1$,
\[
  \begin{aligned}
    &\Bigg|G_{\X_{N(1)}}(t)-G_{\X_1}(t)
      - \sum_{\ell=1}^{k-1}\sum_{\alpha \in \Lambda_\ell}
          \frac{D^\alpha G_{\X_1}(t)}{\alpha!}\,
          (\X_{N(1)}-\X_1)^\alpha\Bigg|
     \lesssim \max_{\beta\in \Lambda_k}\|D^\beta G_{\cdot}(t)\|_{\infty}\,
           \big\|\X_{N(1)}-\X_1\big\|^k,\\
    &\Bigg|\hat{G}_{\X_{N(1)}}(t)-\hat{G}_{\X_1}(t)
      - \sum_{\ell=1}^{k-1}\sum_{\alpha \in \Lambda_\ell}
          \frac{D^\alpha \hat{G}_{\X_1}(t)}{\alpha!}\,
          (\X_{N(1)}-\X_1)^\alpha\Bigg|
     \lesssim \max_{\beta\in \Lambda_k}\|D^\beta \hat{G}_{\cdot}(t)\|_{\infty}\,
           \big\|\X_{N(1)}-\X_1\big\|^k,\\
           &\text{ and }\\
    &\Bigg|\sum_{\ell=1}^{k-1}\sum_{\alpha \in \Lambda_\ell}
              \frac{D^\alpha G_{\X_1}(t)}{\alpha!}\,
              (\X_{N(1)}-\X_1)^\alpha
      - \sum_{\ell=1}^{k-1}\sum_{\alpha \in \Lambda_\ell}
              \frac{D^\alpha \hat{G}_{\X_1}(t)}{\alpha!}\,
              (\X_{N(1)}-\X_1)^\alpha\Bigg|\\
    &\quad\lesssim
      \sum_{\ell=1}^{k-1}
        \max_{\alpha\in \Lambda_\ell}\|D^\alpha G_{\cdot}(t)
                               - D^\alpha \hat{G}_{\cdot}(t)\|_{\infty}\,
        \big\|\X_{N(1)}-\X_1\big\|^\ell.
  \end{aligned}
\]
This gives 
\[
  \begin{aligned}
    \Xi
    &\lesssim  \int
        \Er\bigl[
          \bigl(\max_{\beta\in \Lambda_k}\|D^\beta G_{\cdot}(t)\|_{\infty}
           + \max_{\beta\in \Lambda_k}\|D^\beta \hat{G}_{\cdot}(t)\|_{\infty}\bigr)\,
          \big\|\X_{N(1)}-\X_1\big\|^k\bigr]
        \,\dr\mu(t)\\
    &\quad
      + \sum_{\ell=1}^{k-1}
        \int
          \Er\bigl[
            \max_{\alpha\in \Lambda_\ell}\|D^\alpha G_{\cdot}(t)
                                  - D^\alpha \hat{G}_{\cdot}(t)\|_{\infty}\,
            \big\|\X_{N(1)}-\X_j\big\|^\ell\bigr]
        \,\dr\mu(t)\\
    &\lesssim \Big(\int
            \max_{\beta\in \Lambda_k}\|D^\beta G_{\cdot}(t)\|_{\infty}
          \,\dr\mu(t)\Big)\,
          \Er\bigl[\big\|\X_{N(1)}-\X_1\big\|^k\bigr]\\
    &\quad
      +  \Bigl(\int
                 \Er^{1/2}\bigl[\max_{\beta\in \Lambda_k}\|D^\beta \hat{G}_{\cdot}(t)\|_{\infty}^2\bigr]
                 \,\dr\mu(t)\Bigr)\,
        \Er^{1/2}\bigl[\big\|\X_{N(1)}-\X_1\big\|^{2k}\bigr]\\
    &\quad
      +  \sum_{\ell=1}^{k-1}
        \Bigl(\int
               \Er^{1/2}\bigl[\max_{\alpha\in \Lambda_\ell}\|D^\alpha G_{\cdot}(t)
                                         - D^\alpha \hat{G}_{\cdot}(t)\|_{\infty}^2\bigr]
               \,\dr\mu(t)\Bigr)\,
        \Er^{1/2}\bigl[\big\|\X_{N(1)}-\X_1\big\|^{2\ell}\bigr]\\
    &= o\bigl(n^{-\frac12}\bigr),
  \end{aligned}
\]
where we use again the Cauchy-Swartz inequality in the second inequality, and Assumptions \ref{assum:x}-\ref{assum:estimator} in the final step.

Lastly, we analyze $B_{13}$. It holds true that
\begin{align*}
|B_{13}| &\leq \int \Er[(\hat G_{X_1}(t)-G_{X_1}(t))^2]\dr\mu(t) + \int \Er\Big|(\hat G_{X_1}(t)-G_{X_1}(t))(\hat G_{X_{N(1)}}(t)-G_{X_{N(1)}}(t))\Big|\dr\mu(t)\\
&\leq  2\int\Er\bigl[\big\|G_{\cdot}(t)-\hat{G}_{\cdot}(t)\big\|_{\infty}^2\bigr]
        \,\dr\mu(t)\\
    &= o\bigl(n^{-\frac12}\bigr),
\end{align*}
by Assumption \ref{assum:estimator}. This completes the proof of the first step.

\Step[step2:boundingB2]{Bound $B_2$}
We have 
\[
  \begin{aligned}
    B_2
    &= \int \Er\bigl[\hat{G}_{\X_1}(t)\,\hat{G}_{\X_{N(1)}}(t)
                    - \hat{G}_{\X_1}^2(t)\bigr]
           \,\dr\mu(t)
      - \int \hat{\Er}\bigl[\hat{G}_{\X_1}(t)\,\hat{G}_{\X_{N(1)}}(t)
                    - \hat{G}_{\X_1}^2(t)\bigr]
           \,\dr\mu(t)\\
    &= \int \frac1n
                       \sum_{i=1}^n
                       \lcb
                         \bigl(\Er\big[\hat{G}_{\X_1}(t)\hat{G}_{\X_{N(1)}}(t)
                                      - \hat{G}^2_{\X_1}(t)\bigr]\bigr)
                         - \bigl(\hat{G}_{\X_i}(t)\hat{G}_{\X_{N(i)}}(t)
                                  - \hat{G}^2_{\X_i}(t)\bigr)
                       \rcb
                       \,\dr\mu(t)\\
    &= \frac1n\sum_{i=1}^n (-Z_i)=: U,
  \end{aligned}
\]
where
\[
  Z_i
  = \int \bigl[\hat{G}_{\X_i}(t)\hat{G}_{\X_{N(i)}}(t)-\hat{G}^2_{\X_i}(t)\bigr]\,\dr\mu(t)-\int \Er\bigl[\hat{G}_{\X_1}(t)\hat{G}_{\X_{N(1)}}(t)-\hat{G}^2_{\X_1}(t)\bigr]\,\dr\mu(t).
\]
Now the problem reduces to bounding $U$. Define
\[
  \overline U \coloneqq \frac1n\sum_{i=1}^n (-\overline Z_i),
\]
where
\[
  \overline Z_i
  \coloneqq \int \bigl[G_{\X_i}(t)G_{\X_{N(i)}}(t)-G_{\X_i}^2(t)\bigr]\,\dr\mu(t)-\int \Er\bigl[G_{\X_1}(t)G_{\X_{N(1)}}(t)-G^2_{\X_1}(t)\bigr]\,\dr\mu(t).
\]
Note that $\Er\bigl[\big|U\big|\bigr]
  \le \Er\bigl[\big|\overline U - U\big|\bigr]
      + \Er\bigl[\big|\overline U\big|\bigr]$. Therefore, to show $\Er|B_2|=o(n^{-\frac{1}{2}})$, it suffices to show $\Er\big[|\overline{U}|\big]=o(n^{-\frac{1}{2}})$ and $\Er\bigl[\big|\overline U - U\big|\bigr]=o(n^{-\frac{1}{2}})$.

For $\overline U$, using the McDiarmid's inequality along with Assumption \ref{assum:x}\ref{assum:x1} (leading to the strong convergence of NN distance to 0), Assumption \ref{assum:distribution},  and the fact that each node $i$ can be the NN of at most $O(d)$ many points, changing one input value will only incur an $o_{a.s}(1/n)$ difference in the output. This then yields
\[
  \Er |\overline U|  = o(n^{-1/2}).
\]
Finally, by \cref{step1:boundingB1}, we have 
\[
    \begin{aligned}
        \Er\bigl[\big|U-\overline U\big|\bigr] \le \frac1n\sum_{i=1}^n
      \Er\bigl[\big|\overline Z_i - Z_i\big|\bigr]\lesssim \Xi = o(n^{-\frac12}).
    \end{aligned}
\]
This concludes $\Er |B_2|=o(n^{-\frac{1}{2}})$.

\Step[step3:boundingB3]{Bound $B_3$}

We have
\[
  \begin{aligned}
    B_3
    &= 
         \int \hat{\Er}\bigl[\hat{G}_{\X_1}(t)\hat{G}_{\X_{N(1)}}(t)-\hat{G}_{\X_1}^2(t)\bigr]\dr\mu(t)
       - \frac{1}{n}\sum_{i=1}^n\frac{1}{n-1}\sum_{j\ne i}
          \bigl[\hat{G}_{\X_j}(Y_i)\hat{G}_{\X_{N(j)}}(Y_i)-\hat{G}_{\X_j}^2(Y_i)\bigr]\\
    &= \frac{1}{n}\sum_{i=1}^n(-W_i),
  \end{aligned}
\]
where
\[
  W_i
  := \frac{1}{n-1}\sum_{j\ne i}\Big\{
          \bigl[\hat{G}_{\X_j}(Y_i)\hat{G}_{\X_{N(j)}}(Y_i)-\hat{G}_{\X_j}^2(Y_i)\bigr]
    -  \int \bigl[\hat{G}_{\X_j}(t)\hat{G}_{\X_{N(j)}}(t)-\hat{G}_{\X_j}^2(t)\bigr]\dr\mu(t)\Big\}.
\]
Define
\[
  S \coloneqq \frac{1}{n}\sum_{i=1}^n(-W_i)
  \quad\text{and}\quad
  \overline{S} \coloneqq \frac{1}{n}\sum_{i=1}^n(-\overline{W}_i),
\]
where 
\[
  \overline{W}_i
  \coloneqq \frac{1}{n-1}\sum_{j\ne i}\Big\{
          \bigl[G_{\X_j}(Y_i)G_{\X_{N(j)}}(Y_i)-G_{\X_j}^2(Y_i)\bigr]
    - \int \bigl[G_{\X_j}(t)G_{\X_{N(j)}}(t)-G_{\X_j}^2(t)\bigr]\dr\mu(t)\Big\}.
\]
For analyzing $\overline S$, we introduce $\tilde N(k)= N^{\setminus i}(k)$ to index the NN of $X_k$ in $\{X_j; j \ne i\}$ and 
\[
\tilde S := \frac{1}{n}\sum_{i=1}^n(-\tilde{W}_i)
\]
with
\[
\tilde W_i:=\frac{1}{n-1}\sum_{j\ne i}
          \Big\{\bigl[G_{\X_j}(Y_i)G_{\X_{\tilde N(j)}}(Y_i)-G_{\X_j}^2(Y_i)\bigr]
    - \int \bigl[G_{\X_j}(t)G_{\X_{\tilde N(j)}}(t)-G_{\X_j}^2(t)\bigr]\dr\mu(t)\Big\}.
\]
The new random integer $\tilde N(k)$ has the advantage of being independent of $(X_i,Y_i)$ as long as $k\ne i$, so that $\Er(\tilde W_i)=0$, which gives us $\Er(\tilde S) = 0$. Since $\Er|S|\le \Er|S-\overline{S}|+\Er|\overline{S}-\widetilde{S}|+\Er|\widetilde{S}|$, we analyze the three terms on the right-hand side individually.

{\bf Step 3.1.} We first show $\Er|\widetilde{S}|=o(n^{-\frac{1}{2}})$. For this, we implement a similar argument as in \cref{step2:boundingB2}: using the McDiarmid's inequality along with Assumption \ref{assum:x}\ref{assum:x1}, Assumption \ref{assum:distribution},  and the fact that each node $i$ can be the NN of at most $O(d)$ many points, changing one input value will only incur an $o_{a.s}(1/n)$ difference in the output. Therefore, $\Er|\widetilde{S}|=o(n^{-\frac{1}{2}})$.

{\bf Step 3.2.}  Secondly, we show that $\Er|\overline S-\widetilde S|=O(1/n)$. For this, we have 
\[
|\overline S-\tilde S| \le \frac{4}{n(n-1)}\sum_{i=1}^n \sum_{j\ne i} \Ibbm(N(j)=i)=\frac{4}{n(n-1)}\sum_{j=1}^n \sum_{i\ne j} \Ibbm(N(j)=i)=\frac{4}{n-1}.
\]
Therefore, $\Er|\overline{S}-\widetilde{S}|=O(\frac{1}{n})$.

{\bf Step 3.3.} It remains to relate $S$ to $\overline S$. To this end, similar to \cref{step2:boundingB2}, we can derive
\[
\Er |S-\overline{S}|  = o(n^{-\frac12}),
\]
which shows $ \Er |B_3| = o(n^{-\frac12})$.

This concludes the first statement of the theorem. The second statement is then direct. 
\end{proof}

\begin{proof}[Proof of \cref{thm:uni_approx_rate}]
Proof of this theorem is derived from the following three lemmas. The first one derives an upper bound on $\Er\bigl[\max_{\alpha\in \Lambda_r}\big\|D^\alpha \hat{\psi}_{t,K}
                         - D^\alpha \psi_{t}\big\|_\infty^2\bigr]$ in terms of $\zeta_{0,K}$, $\zeta_{r,K}$, $\underline{\lambda}_K$, $K$, $n$, $\Er\bigl[\big\|\hat{Q}^{-1}\big\|_2\bigr]$, and $\Er\bigl[\big\|\hat{Q}^{-1}\big\|_2^2\,
                \big\|Q^{-\frac12}\,\hat{Q}\,Q^{-\frac12} - I_K\big\|_2^2\bigr]$.
    \begin{lemma}\label{lem:bound_estimator}
    
In the setting of \cref{sec:rls}, we have
\[
  \Er\bigl[\max_{\alpha\in \Lambda_r}\big\|D^\alpha \hat{\psi}_{t,K}
                         - D^\alpha \psi_{t}\big\|_\infty^2\bigr]
  \le \zeta_{r,K}^2\,\underline{\lambda}_K^{-1}
      \bigl(\eta_1 + \eta_2 + \eta_3\bigr)
      + 2\,(\vartheta_{r,K}^t)^2,
\]
where
\[
  \eta_1 = \frac{1}{4}\,\zeta_{0,K}^2\,K\,n^{-1}\,
        \Er\bigl[\big\|\hat{Q}^{-1}\big\|_2\bigr], 
\quad
  \eta_2 = 2\,\zeta_{0,K}^6\,\underline{\lambda}_K^{-1}\,
        \Er\bigl[\big\|\hat{Q}^{-1}\big\|_2^2\,
                \big\|Q^{-\frac12}\,\hat{Q}\,Q^{-\frac12} - I_K\big\|_2^2\bigr],
\quad
  \eta_3 = 2\,\zeta_{0,K}^2\,\underline{\lambda}_K^{-1}\,n^{-1}.
\]
\end{lemma}

The next two lemmas derive orders of $\Er\bigl[\big\|\hat{Q}^{-1}\big\|_2\bigr]$ and $\Er\bigl[\big\|\hat{Q}^{-1}\big\|_2^2\,
                \big\|Q^{-\frac12}\,\hat{Q}\,Q^{-\frac12} - I_K\big\|_2^2\bigr]$ in terms of $\zeta_{0,K}$, $\underline{\lambda}_K$, $K$, $n$, and $\lambda_n$.

\begin{lemma}\label{lem:hatQ}
Let $a$ be a positive integer. Assume $\lambda_n>0$ and $\lambda_n\asymp n^{-c}$ for some $c> 0$. Furthermore, assume 
\[
\zeta_{0,K}=o\Big(\lb\frac{n(\underline{\lambda}_K-\lambda_n)^2}{\log K+\log(\lambda_n^{-a})}\rb^{\frac{1}{4}}\wedge \lb\frac{n(\underline{\lambda}_K-\lambda_n)}{\log K+\log(\lambda_n^{-a})}\rb^{\frac{1}{2}}\Big)
\]
 with $\underline{\lambda}_K>\lambda_n$ for all sufficiently large $n$. Then we have 
 \[
 \Er\big[\big\|\hat{Q}^{-1}\big\|_2^a\big]=O(\underline{\lambda}_K^{-a}) \text{ as } n\to \infty.
 \]   
\end{lemma}

\begin{lemma}\label{lem:hatQ2}
Under \cref{assum:rate}, we have 
  \[
  \Er\big[\big\|\hat{Q}^{-1}\big\|_2^2\big\|Q^{-\frac{1}{2}}\hat{Q}Q^{-\frac{1}{2}}-I_K\big\|_2^2\big] \lesssim \underline{\lambda}_K^{-4}(\underline{\lambda}_K\zeta_{0,K}^2\log(K)n^{-1}+\lambda_n^2).
  \]
\end{lemma}
Note that since $K/n \to 0$ and $\lambda_n\asymp n^{-c}$, we have $\log K+\log (\lambda_n^{-a})\asymp \log n$. In addition, $\zeta_{0,K} = o((n/\log n)^{1/4}(\underline{\lambda}_K - \lambda_n)^{1/2})$ implies $\zeta_{0,K} = o((n/\log n)^{1/2}(\underline{\lambda}_K - \lambda_n)^{1/2})$. Hence, \cref{assum:rate} implies the assumptions of \cref{lem:hatQ}. Combining the three lemmas gives the result.
\end{proof}

\section{Proofs of supporting lemmas}\label{sec:proofs_lem}

\begin{proof}[Proof of \cref{lem:E}]
    We show the first claim. Note that $E_{1,n}=\int\Er[G_{X_1}(t)G_{X_{N(1)}}(t)]\dr \mu(t)$ and $X_{N(1)}$ converges to $X_1$ in probability. Then by the continuity and boundedness of $G$, we have (boundedness of $G$ upgrades convergence in probability to convergence of expectations and allows to interchange limit and integral)
    \[
        E_{1,n}=\int\Er[G_{X_1}(t)G_{X_{N(1)}}(t)]\dr \mu(t) \to \int \Er[G_{X_1}(t)^2]\dr \mu(t).
    \]

    We now show the second claim. Note that 
    \[
        \begin{aligned}
            \Prb(Y_1\le Y_{N(1)}\mid X_1=x,X_{N(1)}=z)&=\Er[\Prb(Y_1\le Y_{N(1)}\mid Y_1,X_1=x,X_{N(1)}=z)]\\
            &=\int \Prb(Y_{N(1)}\ge t\mid X_{N(1)}=z)\Prb(Y_1\in \dr t\mid X_1=x)\\
            &=\int G_z(t) \Prb(Y_1\in \dr t\mid X_1=x),
        \end{aligned}
    \]
    and
    \[
        \Er[G_x(Y_1)\mid X_1=x]=\frac{1}{2}.
    \]
    Therefore, we have
    \[
        \begin{aligned}
            E_{2,n}&=\Er[\Prb(Y_1\le Y_{N(1)}\mid X_1,X_{N(1)})]\\
            &=\Er\Big[\int G_{X_{N(1)}}(t)\Prb(Y_1\in \dr t\mid X_1)\Big]\\
            &\to \Er\Big[\int G_{X_{1}}(t)\Prb(Y_1\in \dr t\mid X_1)\Big]=\frac{1}{2},
        \end{aligned}
    \]
    and the proof is thus complete.
\end{proof}

\begin{proof}[Proof of \cref{lem:Er^2|x}]
    Write $p_r(x)\coloneqq \int_{B(x,r)\cap \Xc}f_X(u)\dr u$. We have uniformly in $x\in \Xc_{r_n}$ and $0<r\le r_n$
    \[
    p_r(x)=f_X(x)v_dr^d+O(r^{d+1}).
    \]
    Set 
    \[
    M_n\coloneqq \widetilde{M}_nn^{-\frac{1}{d}},~~\text{ with } 
    \widetilde{M}_n=\min\Big\{n^{\frac{1}{2d(d+1)}},\frac{1}{2}r_n n^{\frac{1}{d}}\Big\}.
    \] 
    This choice gives $M_n\to 0$ and $nM_n^d\to \infty$.
    Since 
    \[
    M_n\le r_n~~ {\rm and}~~ \log(1-u)=-u-\frac{u^2}{2}+O(u^3), 
    \]
    when $r\le M_n$ we have
    \[
        \log(1-p_r(x))^{n-1}=-(n-1)f_X(x)v_dr^d+O(nr^{d+1})+O(nr^{2d})
    \]
    uniformly in $x\in \Xc_{r_n}$ and $r\le M_n$. Since 
    \[
    nr^{d+1}\le \widetilde{M}_n^{d+1}n^{-\frac{1}{d}}=o(1)~~ {\rm and}~~ nr^{2d}\le \widetilde{M}_n^{2d}n^{-1}=o(1), 
    \]
    we have
    \[
        (1-p_r(x))^{n-1}=\exp(-nf_X(x)v_dr^d)(1+o(1))
    \]
    uniformly in $x\in \Xc_{r_n}$ and $r\in[0,M_n]$. Note that uniformly in $x\in \Xc_{r_n}$
    \[
        \int_{M_n}^\infty r\Prb(R>r\mid X_1=x)\dr r=\int_{M_n}^{r_0}+\int_{r_0}^\Df\lesssim n^{-\frac{2}{d}}e^{-CnM_n^d}+e^{-Cnr_0^d}=o(n^{-2/d}).
    \]
    Hence,
    \begin{align*}
            \Er[R^2\mid X_1=x]&=2\int_0^\infty r\Pr(R>r\mid X_1=x)\dr r\\
            &=2\int_0^{M_n} r(1-p_r(x))^{n-1}\dr r+o(n^{-\frac{2}{d}})\\
            &=2(1+o(1))\int_0^{M_n}r\exp(-nv_df_X(x)r^d)\dr r+o(n^{-\frac{2}{d}})\\
            &=2(1+o(1))\int_0^\infty r\exp(-nv_df_X(x)r^d)\dr r+o(n^{-\frac{2}{d}})\\
            &=\Gamma(1+2/d)(v_df_X(x)n)^{-\frac{2}{d}}+o(n^{-\frac{2}{d}}),
        \end{align*}
     uniformly in $x\in \Xc_{r_n}$.
\end{proof}

\begin{proof}[Proof of \cref{lem:symmetric_difference}]
    Define 
    \[
    \varepsilon_n\coloneqq \sup_{\|h\|\le r_n,|\delta(x)|\le r_n}\frac{|\delta(x+h)-\delta(x)-\nabla\delta(x)\cdot h|}{\|h\|}. 
    \]
    By \cref{ass:C2_boundary}, \citet[Theorem~8.2 in Chapter~7]{delfour2011shapes} implies that function $x\mapsto\delta(x)$ is $C^{1,1}$ locally near the boundary. Since $\partial \Xc$ is compact, we can obtain a finite cover of $\partial \Xc$ so that we obtain a uniform tube radius $r_*>0$ and a uniform Lipschitz constant for $\nabla \delta$ on $\{|\delta|<r_*\}$. Therefore, $\varepsilon_n\downarrow 0$ as $n\to \infty$ and for sufficiently large $n$, it holds for $0\le r\le r_n$
    \[
    \sup_{\|\xi\|=1,|\delta(x)|\le r_n}|\delta(x+r\xi)-(\delta(x)+r\xi\cdot\nabla \delta(x))|\le r\varepsilon_n.
    \] 
    If $\xi\in A_x(r)\setminus C_x(r)$, then 
    \[
    0\le \delta(x+r\xi)\le \delta(x)+r\xi\cdot \nabla \delta(x)+r\varepsilon_n
    \]
     and thus 
     \[
     \delta(x)+r\xi\cdot\nabla \delta(x)\geq -r\varepsilon_n. 
     \]
     If $\xi\in C_x(r)\setminus A_x(r)$, then 
     \[
     0> \delta(x+r\xi)\ge \delta(x)+r\xi\cdot \nabla \delta(x)-r\varepsilon_n 
     \]
     and thus 
     \[
     \delta(x)+r\xi\cdot\nabla \delta(x)< r\varepsilon_n. 
     \]
     Therefore, when $0\le \delta(x)\le r\le r_n$, we have 
     \[
     A_x(r)\sd C_{x}(r)\subseteq M_x(r)\coloneqq\{\xi\in \Sb^{d-1}:|\delta(x)+r\xi\cdot\nabla \delta(x)|\le r\varepsilon_n\}. 
     \]
     Note that when $d\ge 2$
    \[
        \begin{aligned}
            \sup_{0\le \delta(x)\le r\le r_n} \sigma(M_x(r))&=\sup_{0\le \delta(x)\le r\le r_n}\int \Ibbm(y\in M_x(r))\dr \sigma(y)\\
            &\le \sigma(\Sb^{d-2})\sup_{0\le \delta(x)\le r\le r_n}\int_{-\delta(x)/r-\varepsilon_n}^{-\delta(x)/r+\varepsilon_n}(1-u^2)^{\frac{d-3}{2}}\dr u=O(\varepsilon_n\vee \sqrt{\varepsilon_n})=o(1).
        \end{aligned}
    \]

    When $d=1$, $\sigma(A_x(r)\sd C_{x}(r))=0$ for all $0\le \delta(x)<r\le r_n$.
\end{proof}

\begin{proof}[Proof of \cref{lem:ball_asymp}]
    First note that 
    \[
    |B(x,r)\cap \Xc|=r^d\int_{B(0,1)}\Ibbm(x+ru\in \Xc)\dr u=r^d\int_{B(0,1)}\Ibbm(\delta(x+ru)\ge 0)\dr u.
    \]
    Similar to the proof of \cref{lem:symmetric_difference}, we have
    \[
        H_{\tau-\varepsilon_n}\subseteq \{u:\delta(x+ru)\ge 0\}\subseteq H_{\tau+\varepsilon_n},
    \]
    where  $H_{\tau}\coloneqq \{u\in B(0,1):u\cdot \nabla \delta(x)\ge -\tau\}$ for $\tau\coloneqq \delta(x)/r \in [0,1]$.
    Therefore,
    \[
        \{u:\delta(x+ru)\ge 0\}\sd H_{\tau}\subseteq (H_{\tau+\varepsilon_n}\setminus H_{\tau})\cup (H_{\tau}\setminus H_{\tau-\varepsilon_n})\subseteq \widetilde{H}_{\tau},
    \]
    where $\widetilde{H}_{\tau}=\{u:|u\cdot \nabla\delta(x)+\tau|\le \varepsilon_n\}$. Thus,
    \[
        \Big|\int_{B(0,1)}\Ibbm(\delta(x+ru)\ge 0)\dr u-\int_{B(0,1)}\Ibbm(u\in H_{\tau})\dr u\Big|\le |\widetilde{H}_{\tau}|.
    \]
    Write $u=\alpha \nabla\delta(x)+w$ with $w$ orthogonal to $\nabla \delta(x)$ and $\alpha=u\cdot\nabla \delta(x)\in (-1,1)$. Then
    \[  
        \begin{aligned}
            |\widetilde{H}_\tau|&=\int_{\{\alpha:|\alpha+\tau|\le \varepsilon_n\}} |B_{d-1}(0,\sqrt{1-\alpha^2})|\dr \alpha\\
            &= \int_{\{\alpha:|\alpha+\tau|\le \varepsilon_n\}} v_{d-1}(1-\alpha^2)^{\frac{d-1}{2}}\dr \alpha\\
            &\le 2v_{d-1}\varepsilon_n.
        \end{aligned}
    \]
    In addition, note that 
    \[
        \int_{B(0,1)}\Ibbm(u\in H_{\tau})\dr u=|B(0,1)\cap H_\tau|=v_d\frac{\int^1_{-\tau}(1-u^2)^{\frac{d-1}{2}}\dr u}{\int^1_{-1}(1-u^2)^{\frac{d-1}{2}}\dr u}=\frac{1}{2}v_d\Big(1+I_{\tau^2}\Big(\frac{1}{2},\frac{d+1}{2}\Big)\Big).
    \]
    Hence, we have
    \[
    \sup_{(x,r)\in \Sc_n}\Bigg|\frac{|B(x,r)\cap \Xc|-v_d\Ff_d(\delta(x)/r)r^d}{r^d}\Bigg|=o(1)
    \]
    and the proof is thus complete.
\end{proof}

\begin{proof}[Proof of \cref{lem:approximate_bounday_proj}]
    For sufficiently large $n$, define $\Psi:\partial\Xc\times [0,r_n]\to \Xc$ such that 
    \[
    \Psi(y,s)=y+s\n(y), 
    \]
    where $\n(y)$ is the inward normal at $y$. By \cref{lem:normal_offset_map}, $\Psi$ is well defined and a $C^1$-diffeomorphism onto its image $\{x:0\le \delta(x)\le r_n\}$.  Then we have, by the change-of-variable formula,
    \[
        \begin{aligned}
            &n\int_{\{0\le \delta(x)\le r_n\tau\}} \delta(x)^{d+1}\exp(-\rho_x(\tau)\delta(x)^d\tau^{-d})f_X^2(x)\Gf(x)\dr x\\
            =& n\int_{\partial \Xc}\int_0^{r_n\tau} s^{d+1}\exp(-\rho_{y+s\n(y)}(\tau)s^{d}\tau^{-d})f_X^2(y+s\n(y))\Gf(y+s\n(y))J_{\Psi}(y,s)\dr s \dr \Hc^{d-1}(y)\\
            =:& \Upsilon(\tau),
        \end{aligned} 
    \]
    where $J_\Psi$ denotes the Jacobian of $\Psi$. Note that from the conditions on $f_X$ and $G$ and the inequality 
    \[
    e^{-a}-e^{-b}\le e^{-\min\{a,b\}}|a-b| ~~\text{ for }a,b\ge 0, 
    \]
    it holds true, with constants $C$ and $c$ independent of $y,s,\tau$ for all $s,y$ and $\tau\in (0,1]$,
    \[
        \begin{aligned}
            |f_X^2(y+s\n(y))\Gf(y+s\n(y))-f_X^2(y)\Gf(y)|&\le Cs  \quad \text{ and }\\
            |\exp\big(-\rho_{y+s\n(y)}(\tau)s^d\tau^{-d}\big)-\exp\big(-\rho_{y}(\tau)s^d\tau^{-d}\big)|&\le C s^{d+1}\tau^{-d}n\exp(-c\rho_y(\tau)s^d\tau^{-d}).
        \end{aligned}
    \]
    In addition, by \cref{lem:normal_offset_map}, we have 
    \[
    |J_{\Psi}(y,s)-1|\le Cs~~\text{ for }(y,s)\in \partial \Xc\times [0,r_n]. 
    \]
    Therefore, we have
    \[
        \Upsilon(\tau)=n\int_{\partial \Xc}\int_0^{r_n\tau} s^{d+1}\exp(-\rho_{y}(\tau)s^{d}\tau^{-d})f_X^2(y)\Gf(y)\dr s\dr \Hc^{d-1}(y)+\Ec_n(\tau),
    \]
    and by the choice of $r_n$, uniformly in $\tau \in (0,1]$
    \[
        \begin{aligned}
            |\Ec_n(\tau)|&\le Cn\int_{\partial \Xc}\int_0^{r_n\tau} ns^{2d+2}\tau^{-d}\exp\big(-c\rho_y(\tau)s^d\tau^{-d}\big)+s^{d+2}\exp\big(-c\rho_y(\tau)s^d\tau^{-d}\big)\dr s\dr \Hc^{d-1}(y)\\
            &\lesssim n^2r_n^{2d+3}=o(n^{-2/d}).
        \end{aligned}  
    \]
    The remaining task is to evaluate the main term of $\Upsilon(\tau)$. First note that for $a,m>0$, we have
    \[
        \int^r_0 s^m \exp(-as^d)\dr s=a^{-\frac{1+m}{d}}d^{-1}\gamma\Big(\frac{1+m}{d},ar^d\Big),
    \]
    where $\gamma(\cdot,\cdot)$ is the lower incomplete gamma function. Therefore, we have
    \[
        \int_0^{r_n\tau} s^{d+1}\exp(-\rho_y(\tau)\tau^{-d}s^d)\dr s= d^{-1}\big(\rho_y(\tau)\tau^{-d}\big)^{-(1+2/d)}\gamma(1+2/d,\rho_y(\tau)r_n^d).
    \]
    Note that $\gamma(1+2/d,\rho_y(\tau)r_n^d)=\Gamma(1+2/d)+o(1)$ uniformly in $\tau$ and $y$. So the main term of $\Upsilon(\tau)$ is uniformly in $\tau$
    \[
        \Upsilon(\tau)-\Ec_n(\tau)
            =d^{-1}\Gamma(1+2/d)n\int_{\partial \Xc} f_X^2(y) \Gf(y) \big(\rho_y(\tau)\tau^{-d}\big)^{-(1+2/d)}\dr \Hc^{d-1}(y)(1+o(1)).
    \]
    Since  $nf_X(y)\big(\rho_y(\tau)\tau^{-d}\big)^{-(1+2/d)}=n^{-2/d}f_X(y)^{-2/d}\big(v_d\Ff_d(\tau)\tau^{-d}\big)^{-(1+2/d)}$, it holds true that
    \[
        \int_0^1 \Mf_d(\tau)\sigma(C(\tau))\tau^{-(d+2)}n\big(\rho_y(\tau)\tau^{-d}\big)^{-(1+2/d)}\dr \tau\asymp n^{-2/d} \int_0^1 \Mf_d(\tau)\sigma(C(\tau))\Ff_d(\tau)^{-(1+2/d)}\dr \tau,
    \]
    where the integral $\int_0^1 \Mf_d(\tau)\sigma(C(\tau))\Ff_d(\tau)^{-(1+2/d)}\dr \tau$ is finite, and
    \[
        \begin{aligned}
            &\int_{0}^1 \Mf_d(\tau)\sigma(C(\tau))\tau^{-(d+2)} (\Upsilon(\tau)-\Ec_n(\tau))\dr \tau \\
            &=  n^{-2/d}d^{-1}\Gamma(1+2/d)\int_{0}^{1} \Mf_d(\tau)\sigma(C(\tau))\big(v_d\Ff_d(\tau)\big)^{-(1+2/d)}\dr \tau\int_{\partial \Xc}f_X(y)^{1-2/d}\Gf(y)\dr \Hc^{d-1}(y)+o(n^{-2/d}).
        \end{aligned}
    \]

        We now compute the integral 
        \[
        \int_{0}^{1} \Mf_d(\tau)\sigma(C(\tau))\Ff_d(\tau)^{-(1+2/d)}\dr \tau. 
        \]
        By rotation, assume $\nabla \delta(x)=e_d=(0,\ldots,0,1 )\in\Rb^d$. In this case, we can write 
        \[
        C(\tau)=\{\xi\in \Sb^{d-1}:\xi_d\ge -\tau\}. 
        \]
        For $d\ge 2$, write $\xi=(\sqrt{1-u^2}w,u)$ with $u\in[-1,1]$ and $w\in \Sb^{d-2}$. By the symmetry of the first $(d-1)$-coordinates of $C_x(r)$, we have 
    \begin{align*}
    \int_{C(\tau)}\xi \dr \sigma(\xi)&=\int_{-\tau}^1 \int_{\Sb^{d-2}}(\sqrt{1-u^2}w,u)(1-u^2)^{\frac{d-3}{2}}\dr \sigma_{d-2}(w)\dr u\\
    &=\sigma_{d-2}(\Sb^{d-2})\Big(\int_{-\tau}^1 u(1-u^2)^{\frac{d-3}{2}}\dr u\Big)e_d,
    \end{align*}
    which implies $\Mf_d(\tau)\sigma(C(\tau))=v_{d-1}(1-\tau^2)^{\frac{d-1}{2}}$.
    Also note that $\Ff_d^\prime(\tau)=v_{d-1}v_d^{-1}(1-\tau^2)^{\frac{d-1}{2}}$. Therefore, $\Mf_d(\tau)\sigma(C(\tau))=v_d\Ff_d^\prime (\tau)$. When $d=1$, after rotation, we have $C_x(r)=\{\xi \in \{-1,1\}:\xi \ge -\tau\}$. Note that
    \[
            \int_{C_x(r)}\xi \dr \sigma=\sum_{\xi\in C_x(r)} \xi =\Ibbm(0\le \tau <1)\quad \text{ and }\quad 
            \int_{C_x(r)}\ \dr \sigma=\#C_x(r)=2-\Ibbm(0\le \tau <1).
    \]
    Hence, $\Mf_1(\tau)\sigma(C(\tau))=\Ibbm(0\le \tau<1)$. Also note that $\Ff_1(\tau)=\frac{1}{2}(1+\tau)$ and $v_1\Ff_1^\prime(\tau)=1$ for $\tau\in (0,1)$. Therefore, for $\tau\in (0,1)$, it holds $\Mf_1(\tau)\sigma(C(\tau))=v_1\Ff_1^\prime(\tau)$. Hence, for $d\ge 1$
    \[
        \int_{0}^{1} \Mf_d(\tau)\sigma(C(\tau))\Ff_d(\tau)^{-(1+2/d)}\dr \tau=v_d\int_0^1 \Ff_d^\prime(\tau)\Ff_d(\tau)^{-(1+2/d)}\dr \tau=\frac{dv_d}{2}(2^{2/d}-1),
    \]
    where we use the facts that 
    \[
    \frac{\dr}{\dr \tau}\Ff_d(\tau)^{-2/d}=-\frac{2}{d}\Ff_d^\prime(\tau)\Ff_d(\tau)^{-(1+2/d)}~~~{\rm and}~~~ \Ff_d(1)=1 \text{ and }\Ff_d(0)=\frac{1}{2}. 
    \]
    Overall, we have
    \[
        \varpi=\int_{0}^1 \Mf_d(\tau)\sigma(C(\tau))\tau^{-(d+2)} \Upsilon(\tau)\dr \tau=\widetilde{\Cf}_3 n^{-2/d}+o(n^{-2/d}),
    \]
    where
    \[
        \widetilde{\Cf}_3=\frac{1}{2}(2^{2/d}-1)v_d^{-2/d}\Gamma(1+2/d)\int_{\partial \Xc} f_X(y)^{1-2/d}\Gf(y)\dr \Hc^{d-1}(y).
    \]
This completes the proof.
\end{proof}

\begin{lemma}\label{lem:Lip_boundary_interior_nondegeneracy}
    Suppose \cref{assum:ass_4}. Then there exist constants $r_0>0$ and $c_0\in (0,1)$ such that for every $x\in \Xc$ and $0\le r\le r_0$, $|\Xc\cap B(x,r)|\ge c_0v_dr^d$.
\end{lemma}

\begin{proof}[Proof of \cref{lem:Lip_boundary_interior_nondegeneracy}]
    By \citet[Theorem~1.2.2.2]{Grisvard11Elliptic_nonsmooth_domains}, there exist $\theta\in(0,\frac{\pi}{2}]$ and $h^*>0$ such that for every $y\in \partial\Xc$ there is a new coordinate for which  $y-\Cc_{\theta,h^*}\subseteq \Xc$ where $\Cc_{\theta,h^*}\coloneqq \{(z^\prime,z_d):(\cot\theta)\|z^\prime\|<z_d<h^*\}$. 
    
    Fix $y\in \partial \Xc$. Let $0<\rho\le r_0\coloneqq \min\{\frac{h^*}{2},1\}$. Set $s\coloneqq \frac{\rho}{2}$ and take the point $z\coloneqq y-(0,\ldots,0,s)$.  We show for $\kappa\coloneqq \frac{1}{4}\min \{1,\frac{1}{1+\cot\theta}\}$, it holds 
    \begin{equation}\label{eq:ball_inclusion}
        B(z,\kappa \rho)\subseteq (y-\Cc_{\theta,h^*})\cap B(y,\rho)\subseteq \Xc\cap B(y,\rho).
    \end{equation}
    For any $w\in B(z,\kappa \rho)$, we can write $w=y-(w^\prime,w_d)$ with $\|(w^\prime,w_d)-(0,s)\|<\kappa \rho$. Then $\|w^\prime\|<\kappa \rho$ and $|w_d-s|<\kappa \rho$. Therefore, by the choice of $\kappa$
    \[
        (\cot\theta)\|w^\prime
        \|< (\cot\theta)\kappa \rho<s-\kappa \rho< w_d.
    \]
    Since $w_d<\kappa \rho+s$ and $\kappa \rho\le s\le \frac{h^*}{2}$, it yields $w_d<h^*$. Hence, $w$ satisfies the cone inequality and $w\in y-\Cc_{\theta,h^*}$. Note that $\|w-y\|\le \|z-y\|+\kappa \rho=s+\kappa \rho< \rho$. So $w\in B(y,\rho)$.

    Now take arbitrary $x\in \Xc$ and $0<r\le r_0$. If $\delta(x)\ge \frac{r}{2}$, then $B(x,\frac{r}{2})\subseteq \Xc$ and $|B(x,r)\cap \Xc|\ge 2^{-d}v_dr^d$. Otherwise, pick $y\in \partial \Xc$ with $\|x-y\|=\delta(x)<\frac{r}{2}$. Apply \cref{eq:ball_inclusion} with $\rho=\frac{r}{2}$ to get
    \[
        B\big(z,\frac{\kappa}{2}r\big)\subseteq \Xc\cap B\big(y,\frac{r}{2}\big)\subseteq \Xc\cap B(x,r).
    \]
    Hence, $|B(x,r)\cap \Xc|\ge \big(\frac{\kappa}{2}\big)^d v_dr^d$. Setting $c_0\coloneqq \min\{2^{-d},\big(\frac{\kappa}{2}\big)^d\}$ gives $|\Xc\cap B(x,r)|\ge c_0v_dr^d$.
\end{proof}

\begin{lemma}\label{lem:Lip_boundary_linear_bound}
    Suppose \cref{assum:ass_4}. Then there exists a constant $C>0$ such that for all $0\le r\le r_0$
    \[
        \Prb(\delta(X_1)\le r)\le \|f_X\|_{\infty}\big|\{x\in \Xc: \delta(x)\le r\}\big|\le Cr.
    \]
\end{lemma}

\begin{proof}[Proof of \cref{lem:Lip_boundary_linear_bound}]
    Since $\Xc$ is compact and has Lipschitz boundary, up to a rigid motion, there exist finitely many open sets $U_i=V_i\times (a_i,b_i)$ with open set $V_i\subseteq \Rb^{d-1}$, covering $\partial \Xc$ such that 
    \[
        \Xc\cap U_i=\{(v,t)\in V_i\times (a_i,b_i): t\ge \varphi_i(v)\},
    \]
    where $\varphi_i:V_i\to \Rb$ are $\Lc$-Lipschitz such that $\varphi_i(V_i)\subseteq (a_i+\varepsilon,b_i-\varepsilon)$ for some small $\varepsilon>0$. 
    When $r\le r_*$ is small enough, since $\varphi_i$ is $\Lc$-Lipschitz, it holds for some constant $c_\Lc$
    \[
        \{x\in \Xc\cap U_i:\delta(x)\le r\}\subseteq \{(v,t):v\in V_i, \varphi_i(v)\le t\le \varphi_i(v)+c_\Lc r\}.
    \]
    Indeed, take a point $x=(v,\varphi_i(v)+s)\in \Xc\cap U_i$ with height $s>0$ above the boundary graph. For any boundary point $z=(v^\prime,\varphi_i(v^\prime))\in \partial \Xc\cap U_i$,
    \[  
        \begin{aligned}
            \|x-z\|^2&=\|v-v^\prime\|^2+(\varphi_i(v)+s-\varphi_i(v^\prime))^2\\
            &\ge \|v-v^\prime\|^2+(s-|\varphi_i(v)-\varphi_i(v^\prime)|)^2\\
            &\ge \|v-v^\prime\|^2+(s-\Lc\|v-v^\prime\|)^2\\
            & =(1+\Lc^2)\|v-v^\prime\|^2-2s \Lc\|v-v^\prime\|+s^2\eqcolon g(\|v-v^\prime\|).
        \end{aligned}
    \]
    The quadratic function $g$ has minimum value $\frac{s^2}{1+\Lc^2}$. Therefore,
    \[
        \delta(x)\ge \frac{s}{\sqrt{1+\Lc^2}}=: \frac{s}{c_\Lc}.
    \]
    To ensure the vertical strip stays inside the cylinder $V_i\times (a_i,b_i)$ for all $v\in V_i$, pick $0<r_*< \frac{\varepsilon}{c_\Lc}$. Therefore, if $\delta(x)\le r\le r_*$, then we have $0<s\le c_\Lc r$, which is equivalent to
    \[
        \Big\{x\in \Xc\cap U_i:\delta(x)\le r\Big\}\subseteq \Big\{(v,t):v\in V_i, \varphi_i(v) \le t\le \varphi_i(v)+c_\Lc r\Big\}.
    \]
    We then have the following Euclidean volume bound
    \[
        |\{x\in \Xc\cap U_i:\delta(x)\le r\}|\le \int_{V_i}\int_{\varphi_i(v)}^{\varphi_i(v)+c_Lr} \dr t\dr v\le  c_\Lc|V_i|r.
    \]
    Thus 
    \[
        |\{x\in \Xc:\delta(x)\le r\}|\le c_{\Lc}\sum_i\lambda_{d-1}(V_i) r\le Cr.
    \]
    Hence, 
    \[
        \Prb(\delta(X)\le r)=\int_{\delta\le r}f_X(x)\dr x\le \|f_X\|_{\infty}|\{\delta\le r\}|\le Cr.
    \]
    For $r_*<r\le r_0$, we use the trivial bound $\Prb(\delta(X)\le r)\le 1\le \frac{1}{r_*}r$. Overall, for all $0\le r\le r_0$, we obtain the linear bound 
    \[
        \Prb(\delta(X)\le r)\le Cr
    \]
    and thus complete the proof of this lemma.
\end{proof}

\begin{lemma}\label{lem:boundary_mea}
    Let $\Xc\subseteq \Rb^d$ be compact with Lipschitz boundary $\partial \Xc$. Fix $x\in \Rb^d$. For $r>0$, define the sphere $S_r(x)\coloneqq \{u\in \Rb^d:\|u-x\|=r\}=\partial B(x,r)$ and the set of directions $\Pi_r(x)\coloneqq\{\xi\in \Sb^{d-1}:x+r\xi \in \partial \Xc\}$. Then 
    \[
        \Hc^{d-1}(\partial \Xc \cap S_r(x))=0 \quad \text{ and } \quad  \sigma(\Pi_r(x))=0 \quad \text{ for a.e.\ } r>0.
    \]
\end{lemma}

\begin{proof}[Proof of \cref{lem:boundary_mea}]
    Since $\Xc$ is compact and has Lipschitz boundary, $\Hc^d(\partial \Xc)=0$. Define $\varphi:\Rb^d\to \Rb$, $\varphi(u)\coloneqq \|u-x\|$. It is $1$-Lipschitz with $|\nabla \varphi(u)|=1$ for all $u\ne x$. By the coarea formula \cite[Theorem~3.10]{evans2015measure}, we have
    \[
       0=\Hc^d(\partial \Xc)= \int_{\partial \Xc} \dr u=\int_0^\infty \Hc^{d-1}(\partial \Xc \cap S_r(x)) \dr r.    
    \]
    Since $\Hc^{d-1}(\partial \Xc \cap S_r(x))$ is nonnegative, it holds $\Hc^{d-1}(\partial \Xc \cap S_r(x))=0$ for a.e.\ $r>0$.

    The map $\Psi_r:\Sb^{d-1}\to S_r(x)$, defined as $\Psi_r(\xi)=x+r\xi$, is a $C^\infty$-diffeomorphism with Jacobian $r^{d-1}$. Thus, $\Psi_r$ pushforwards surface measure $\sigma$ on $\Sb^{d-1}$  to  $r^{-(d-1)}\Hc^{d-1}$ on $S_r(x)$. Hence, for a.e.\ $r>0$, we have
    \[
        \sigma(\Pi_r(x))=\sigma(\{\xi:x+r\xi\in \partial \Xc\})=r^{-(d-1)}\Hc^{d-1}(\partial \Xc \cap S_r(x))=0
    \]
    and thus complete the proof.
\end{proof}

\begin{lemma}[Conditional density]\label{lem:cond_density}
    For $n\ge 2$, let $X_1,\ldots,X_n$ be i.i.d.\ in $\Rb^d$ with density $f_X$ and support $\Xc\subseteq \Rb^d$. Assume that $\Xc$ has Lipschitz boundary and there exists a continuous function $f\in C(\Rb^d)$ such that $f=f_X$ a.e.\ on $\Xc$.  Fix $x\in \Xc$ and for $r\ge 0$ define
    \[
        p_r(x)\coloneqq\int_{B(x,r)\cap \Xc} f_X(t)\dr t, \quad A_x(r)\coloneqq \Big\{\xi \in \Sb^{d-1}:x+r\xi \in \Xc\Big\}
    \]
    and 
    \[
        R\coloneqq \min_{2\le j\le n}\|X_j-X_1\|, \quad \Xi\coloneqq \frac{X_{N(1)}-X_1}{\|X_{N(1)}-X_1\|}\in \Sb^{d-1}.
    \]
   Then we have:
    \begin{enumerate}[itemsep=0pt,label=(\roman*)]  
        \item the joint conditional density of $(R,\Xi)$ given $X_1=x$ with respect to $\dr r\dr \sigma(\xi)$, is
        \[
            g_{x}(r,\xi)=(n-1)r^{d-1}(1-p_r(x))^{n-2}f_X(x+r\xi)\Ibbm(\xi \in A_x(r));
        \]
        \item the conditional density of $\Xi$ given $R=r$ and $X_1=x$ with respect to $\sigma$ is
        \[
            \pi_{r,x}(\xi)=\frac{f_X(x+r\xi)\Ibbm(\xi\in A_x(r))}{\int_{A_x(r)} f_X(x+r\zeta)\dr \sigma(\zeta)}.
        \]
    \end{enumerate}
\end{lemma}

\begin{proof}[Proof of \cref{lem:cond_density}]
    For Borel set $A\subseteq \Sb^{d-1}$ and $ \varepsilon >0$, define 
    \[
    S_{r, \varepsilon}(x;A)\coloneqq \{x+s\xi:r\le s<r+ \varepsilon,\xi \in A\}. 
    \]
    Set
    \[
        \begin{aligned}
            q_{r, \varepsilon}(x;A)\coloneqq \Prb(X_2\in S_{r, \varepsilon}(x;A)\cap \Xc)&=\int_{S_{r,\varepsilon}(x;A)\cap \Xc} f_X(t)\dr t\\
            &=\int_{r}^{r+ \varepsilon} s^{d-1} \int_A h_s(\xi) \dr \sigma(\xi)\dr s,
        \end{aligned}
    \]
    where 
    \[
    h_s(\xi):=f(x+s\xi)\Ibbm(x+s\xi \in \Xc). 
    \]
    Note that $h_s(\xi) \to h_r(\xi)$ as $s\downarrow r$ pointwise for every $\xi$ such that $x+r\xi \in \Xc\setminus \partial \Xc$. \cref{lem:boundary_mea} implies 
    \[
    \sigma(\{\xi\in \Sb^{d-1}:x+r\xi \in \partial \Xc\})=0 ~~\text{ for a.e. } r>0. 
    \]
    So, $h_s\to h_r$ as $s\downarrow r$ $\sigma$-a.e.\ for a.e.\ $r>0$. Fix a small $ \varepsilon_0>0$ and consider 
    \[
    \Kc\coloneqq \{x+s\xi:s\in [r,r+ \varepsilon_0],\xi\in \Sb^{d-1}\}. 
    \]
    Then $\Kc\cap \Xc$ is compact. By the continuity of $f$, we have 
    \[
    \sup_{x\in \Kc\cap \Xc}f(x)<\infty. 
    \]
    Note that for every $s\in [r,r+ \varepsilon_0]$, the function $h_s$ is dominated by $\sup_{x\in \Kc\cap \Xc}f(x)$ and 
    \[
    \frac{1}{ \varepsilon}\int_r^{r+ \varepsilon}s^{d-1}\dr s\to r^{d-1}~~\text{ as } \varepsilon \downarrow 0. 
    \]
    Then, the dominated convergence theorem gives
    \[
        \frac{q_{r, \varepsilon}(x;A)}{ \varepsilon}\to r^{d-1}\int_{A\cap A_x(r)}f_X(x+r\xi) \dr \sigma(\xi) \quad \text{ as } \varepsilon\downarrow 0.
    \]
    It means that
    \[
        q_{r, \varepsilon}(x;A)= \varepsilon r^{d-1}\int_{A\cap A_x(r)}f_X(x+r\xi) \dr \sigma(\xi)+o( \varepsilon) \quad \text{ as } \varepsilon\downarrow 0.
    \]
    Set 
    \[
    \chi_{r}\coloneqq \sum_{j=2}^n \Ibbm(X_j\in B(x,r)\cap \Xc)~~~{\rm and }~~~\chi^\prime_{r, \varepsilon}(A)\coloneqq \sum_{j=2}^n \Ibbm(X_j\in S_{r, \varepsilon}(x;A)\cap \Xc). 
    \]
    We then have, when $ \varepsilon>0$ is sufficiently small,
    \[
        \begin{aligned}
            \Prb(\Xi\in A, R\in [r,r+ \varepsilon)\mid X_1=x)&=\Prb\Big(\chi_r=0,\chi_{r, \varepsilon}^\prime(\Sb^{d-1})\ge 1, \frac{X_{N(1)}-x}{\|X_{N(1)}-x\|}\in A\Big)\\
            &= \Prb\Big(\chi_r=0,\chi_{r, \varepsilon}^\prime(\Sb^{d-1})= 1, \frac{X_{N(1)}-x}{\|X_{N(1)}-x\|}\in A\Big)+O( \varepsilon^2)\\
            &= \sum_{j=2}^n \Prb(X_j\in S_{r, \varepsilon}(x;A),[X_k\notin B(x,r)\cup S_{r, \varepsilon}(x;\Sb^{d-1})]_{k\ne j})+O( \varepsilon^2)\\
            &= (n-1)q_{r, \varepsilon}(x;A)(1-p_r(x)-q_{r, \varepsilon}(x))^{n-2}+O( \varepsilon^2)\\
            &= (n-1)q_{r, \varepsilon}(x;A)(1-p_r(x))^{n-2}+O( \varepsilon^2)
        \end{aligned}
    \]
    since $\Prb(\chi_{r, \varepsilon}^\prime(\Sb^{d-1})\ge  2)=\binom{n-1}{2}q_{r, \varepsilon}(x;\Sb^{d-1})^2=O( \varepsilon^2)$ and $X_1,\ldots,X_n$ are i.i.d.. Dividing both sides by $ \varepsilon$ and sending $ \varepsilon \downarrow 0$ give the joint density of $(R,\Xi)$ given $X_1=x$, with respect to $\dr r\dr \sigma(\xi)$, which is
        \[
            g_{x}(r,\xi)=(n-1)r^{d-1}(1-p_r(x))^{n-2}f_X(x+r\xi)\Ibbm(\xi \in A_x(r)).
        \]
   The second claim is obvious from the first one.
\end{proof}

\begin{lemma}\label{lem:integral}
    Let $a\ge 0$ be an integer. Then for any $c\in (0,1)$
    \[
        \int_{x}^\infty r^ae^{-nr^d}\dr r\lesssim n^{-\frac{a+1}{d}}e^{-cnx^{d}}.
    \]
\end{lemma}

\begin{proof}[Proof of \cref{lem:integral}]
    Note that
    \[
        \begin{aligned}
            \int_{x}^\infty r^a e^{-nr^d}\dr r &= \int_{nx^d}^\infty \Big(\frac{t}{n}\Big)^{\frac{a}{d}} e^{-t}d^{-1}n^{-\frac{1}{d}}t^{\frac{1}{d}-1}\dr t \qquad  && (t\coloneqq nr^d)\\ 
            &=  d^{-1}n^{-\frac{a+1}{d}} \int_{nx^d}^\infty t^{\frac{1+a}{d}-1}e^{-t}\dr t\\
            &= d^{-1}n^{-\frac{a+1}{d}} \Gamma\Big(\frac{1+a}{d},nx^d\Big).
        \end{aligned}
    \]
    We now analyze the Gamma function: for $0<\varepsilon<1$
    \[
        \Gamma(s,x)=\int_x^\infty t^{s-1}e^{-t}\dr t\le C \int_x^\infty e^{-(1-\varepsilon)t}\dr t=\frac{C}{1-\varepsilon} e^{-(1-\varepsilon)x}.
    \]
    Therefore, $ \int_{x}^\infty r^ae^{-nr^d}\dr r\lesssim n^{-\frac{a+1}{d}}e^{-cnx^{d}}$.
\end{proof}

\begin{lemma}\label{lem:normal_offset_map}
     Let $\Xc\subseteq \Rb^d$ be compact with $C^2$-boundary. There then exists a small $\check{r}>0$, such that the map $\Psi:\partial\Xc\times [0,\check{r}]\to \Xc$, 
     \[
     \Psi(y,s)=y+s\n(y), 
     \]
     where $\n(y)$ is the inward normal at $y$, is a $C^1$-diffeomorphism onto its image $\{x\in\Xc:0\le \delta(x)\le \check r\}$. Furthermore, the Jacobian $J_{\Psi}$ of $\Psi$ satisfies 
     \[
     |J_{\Psi}(y,s)-1|\le Cs,~~\text{ for all }(y,s)\in \partial \Xc\times [0,\check{r}] 
     \]
     with some constant $C>0$.
\end{lemma}

\begin{proof}[Proof of \cref{lem:normal_offset_map}]
From \citet[Theorems~7.2, 7.3, and 8.2 in Chapter~7]{delfour2011shapes}, there exists $\check{r}$ such that the map $\Psi$ is well defined and is a $C^1$-diffeomorphism onto $\{x\in\Xc:0\le \delta(x)\le \check r\}$. By \citet[Theorem~8.5 in Chapter~7]{delfour2011shapes} and smoothness of the determinant map, for some constant $C>0$ and all $(y,s)\in \partial \Xc\times [0,\check{r}]$, it holds true that 
\[
|J_{\Psi}(y,s)-1|\le Cs
\]
and the proof is thus complete.
\end{proof}

\begin{proof}[Proof of \cref{lem:bound_estimator}]

First note that
\[
  \max_{\alpha\in \Lambda_r}
    \big\|D^\alpha p_K^\top\bigl(\hat{\beta}_{t,K}-\beta_{t,K}\bigr)\big\|_\infty^2
  \le \zeta_{r,K}^2\,
      \|\hat{\beta}_{t,K}-\beta_{t,K}\|^2.
\]
We also have
\[
  \|\hat{\beta}_{t,K}-\beta_{t,K}\|^2
  \le \underline{\lambda}_K^{-1}\,
    \|\hat{\psi}_{t,K}-\psi_{t,K}\|_{L^2}^2,
\]
since
\[
  \|\hat{\psi}_{t,K}-\psi_{t,K}\|_{L^2}^2
  = \bigl(\hat{\beta}_{t,K}-\beta_{t,K}\bigr)^\top
    Q\,
    \bigl(\hat{\beta}_{t,K}-\beta_{t,K}\bigr).
\]
Therefore,
\[
  \begin{aligned}
    &\Er\bigl[\max_{\alpha\in \Lambda_r}
               \big\|D^\alpha \hat{\psi}_{t,K}
                        - D^\alpha \psi_{t}\big\|_\infty^2\bigr]\\
    \quad\le& 2\,
      \Er\bigl[\max_{\alpha\in \Lambda_r}
               \big\|D^\alpha p_K^\top
                        \bigl(\hat{\beta}_{t,K}-\beta_{t,K}\bigr)\big\|_\infty^2\bigr]
      + 2\,\max_{\alpha\in \Lambda_r}
          \|D^\alpha\psi_{t,K}-D^\alpha\psi_t\|_{\infty}^2\\
    \quad\le& 2\,\zeta_{r,K}^2\,\underline{\lambda}_K^{-1}\,
            \Er\bigl[\|\hat{\psi}_{t,K}-\psi_{t,K}\|_{L^2}^2\bigr]
          + 2\,(\vartheta_{r,K}^t)^2.
  \end{aligned}
\]
So now we analyze \(\Er[\|\hat{\psi}_{t,K}-\psi_{t,K}\|_{L^2}^2]\). We can bound $\|\hat{\psi}_{t,K}-\psi_{t,K}\|_{L^2}^2$ by two terms as follows:
\[
  \begin{aligned}
    \|\hat{\psi}_{t,K}-\psi_{t,K}\|_{L^2}^2
    &= \int 
       \bigl(p_K(\x)^\top(\hat{\beta}_{t,K}-\beta_{t,K})\bigr)^2
       \,\dr F_{\X}(\x)\\
    &= \bigl(\hat{\beta}_{t,K}-\beta_{t,K}\bigr)^\top
       \Er\bigl[p_K(\X)p_K(\X)^\top\bigr]
       \bigl(\hat{\beta}_{t,K}-\beta_{t,K}\bigr)\\
    &= \big\|Q^{\frac12}\,
             (\hat{\beta}_{t,K}-\beta_{t,K})\big\|^2\\
    &= \big\|Q^{\frac12}\,
             \bigl((P^\top P + n\lambda_n I_K)^{-1}P^\top \Ibbm(Y_{[n]}\ge t)
                  - \beta_{t,K}\bigr)\big\|^2\\
    &\le 2\,
      \big\|Q^{\frac12}\,(P^\top P + n\lambda_n I_K)^{-1}P^\top \varepsilon_t\big\|^2
      + 2\,
      \big\|Q^{\frac12}\,
             \bigl((P^\top P + n\lambda_n I_K)^{-1}P^\top \Psi_t
                  - \beta_{t,K}\bigr)\big\|^2\\
    &= 2\,
      \big\|Q^{\frac12}\,\hat{Q}^{-1}P^\top \varepsilon_t / n\big\|^2
      + 2\,
      \big\|Q^{\frac12}\,
             \bigl(\hat{Q}^{-1}P^\top \Psi_t / n
                  - \beta_{t,K}\bigr)\big\|^2,
  \end{aligned}
\]
where \(\hat{Q} = (P^\top P + n\lambda_n I_K)/n\), 
\(\varepsilon_t = \Ibbm(Y_{[n]}\ge t) - \Psi_t\), and 
\(\Psi_t = (\psi_t(\X_1),\ldots,\psi_t(\X_n))^\top\). In the following, we analyze these two terms individually.

\Step[step1:firstterm]{Analyze the first term}
For the first term, we have
\[
  \begin{aligned}
    \big\|Q^{\frac12}\,\hat{Q}^{-1}P^\top \varepsilon_t / n\big\|^2
    &\le \big\|Q^{\frac12}\,\hat{Q}^{-\frac12}\big\|_2^2\,
      \big\|\hat{Q}^{-\frac12}P^\top \varepsilon_t\big\|^2 / n^2\\
    &= \big\|Q^{\frac12}\,\hat{Q}^{-1}\,Q^{\frac12}\big\|_2\,
      \big\|\hat{Q}^{-\frac12}P^\top \varepsilon_t\big\|^2 / n^2.
  \end{aligned}
\]
We have
\[
  \frac{\Er\bigl[\big\|\hat{Q}^{-\frac12}P^\top \varepsilon_t\big\|^2 \mid \Fc_n\bigr]}{n^2}
  = \frac{\Tr\bigl(\hat{Q}^{-\frac12}P^\top
                   \Er[\varepsilon_t\varepsilon_t^\top \mid \Fc_n]
                   P\,\hat{Q}^{-\frac12}\bigr)}{n^2},
\]
where \(\Fc_n\coloneqq \sigma(\X_1,\ldots,\X_n)\) is the $\sigma$-algebra generated by $X_1,\ldots,X_n$.
For each \(i\), given \(\Fc_n\), the random variable \(\Ibbm(Y_i\ge t)\) is Bernoulli with parameter \(\psi_t(\X_i)\) and therefore
\[
  \Er[\varepsilon_{t,i} \mid \Fc_n] = 0,
  \quad
  \Var(\varepsilon_{t,i}\mid \Fc_n) = \psi_t(\X_i)\bigl(1 - \psi_t(\X_i)\bigr) \le \frac14.
\]
Note that \(Y_i\) is independent of \(Y_j\) given \(\Fc_n\) for \(i\ne j\). Hence,
\[
  \Er[\varepsilon_t\varepsilon_t^\top \mid \Fc_n]
  = \Diag\bigl(\Var(\varepsilon_{t,1}\mid \Fc_n),\ldots,\Var(\varepsilon_{t,n}\mid \Fc_n)\bigr)
  \preceq \frac14\,I_n.
\]
It follows that
\[
  \frac{\Tr\bigl(\hat{Q}^{-\frac12}P^\top
                   \Er[\varepsilon_t\varepsilon_t^\top \mid \Fc_n]
                   P\,\hat{Q}^{-\frac12}\bigr)}{n^2}
  \le \frac14\,\frac{\Tr(A^\top A)}{n^2},
  \quad
  A \coloneqq P\,\hat{Q}^{-\frac12}.
\]
Since \(P^\top P = n(\hat{Q}-\lambda_n I_K)\), we have
\[
  A^\top A = n\,(I_K - \lambda_n \hat{Q}^{-1})
  \preceq n\,I_K, \quad \text{ and }
  \quad
  \Tr(A^\top A)\le n\,\Tr(I_K)=nK.
\]
Thus
\[
  \frac{\Er\bigl[\big\|\hat{Q}^{-\frac12}P^\top \varepsilon_t\big\|^2 \mid \Fc_n\bigr]}{n^2}
  \le \frac{K}{4n}.
\]
Therefore,
\[
  \begin{aligned}
    \frac{\Er\bigl[\|Q^{\frac12}\,\hat{Q}^{-1}\,Q^{\frac12}\|_2\,
                   \big\|\hat{Q}^{-\frac12}P^\top \varepsilon_t\big\|^2\bigr]}{n^2}&= \frac{\Er\bigl[\|Q^{\frac12}\,\hat{Q}^{-1}\,Q^{\frac12}\|_2\,
                     \Er\bigl[\big\|\hat{Q}^{-\frac12}P^\top \varepsilon_t\big\|^2 \mid \Fc_n\bigr]\bigr]}{n^2}\\
    &\le \Er\bigl[\|Q^{\frac12}\,\hat{Q}^{-1}\,Q^{\frac12}\|_2\bigr]\;\frac{K}{4n}\\
    &\le \|Q\|_2\;\Er\bigl[\|\hat{Q}^{-1}\|_2\bigr]\;\frac{K}{4n}\\
    & \le \zeta_{0,K}^2\;\Er\bigl[\|\hat{Q}^{-1}\|_2\bigr]\;\frac{K}{4n}.
  \end{aligned}
\]

\Step[step1:firstterm]{Analyze the second term}

Now consider the second term:
\[
  \begin{aligned}
    &\Er\Bigl[\big\|Q^{\frac12}\bigl(\hat{Q}^{-1}P^\top \Psi_t/n
                           - \beta_{t,K}\bigr)\big\|^2\Bigr]\\
    \quad=& \Er\Bigl[\big\|Q^{\frac12}\bigl(\hat{Q}^{-1}P^\top \Psi_t/n
                            - Q^{-1}P^\top \Psi_t/n
                            + Q^{-1}P^\top \Psi_t/n
                            - Q^{-1}\Er[p_K(\X)\psi_t(\X)]\bigr)\big\|^2\Bigr]\\
    \quad\le& 2\,\Er\Bigl[\big\|Q^{\frac12}(\hat{Q}^{-1}-Q^{-1})P^\top \Psi_t/n\big\|^2\Bigr]
        +2\,\Er\Bigl[\big\|Q^{-\frac12}\bigl(P^\top \Psi_t/n
                           - \Er[p_K(\X)\psi_t(\X)]\bigr)\big\|^2\Bigr].
  \end{aligned}
\]
For the first piece, note \(\frac1nP^\top\Psi_t = \frac1n\sum_i p_K(\X_i)\,\psi_t(\X_i)\) and
\(\big\|\frac1n\sum_i p_K(\X_i)\psi_t(\X_i)\big\|\le \zeta_{0,K}\).  Then
\[
  \begin{aligned}
    \Er\Bigl[\big\|Q^{\frac12}(\hat{Q}^{-1}-Q^{-1})P^\top \Psi_t/n\big\|^2\Bigr]&\le \Er\Bigl[\big\|Q^{\frac12}(\hat{Q}^{-1}-Q^{-1})\big\|_2^2\,
                      \big\|P^\top \Psi_t/n\big\|^2\Bigr]\\
    &\le \zeta_{0,K}^2\,
      \Er\Bigl[\big\|Q^{\frac12}(\hat{Q}^{-1}-Q^{-1})\big\|_2^2\Bigr]\\
    &= \zeta_{0,K}^2\,
      \Er\Bigl[\big\|Q^{\frac12}(\hat{Q}^{-1}-Q^{-1})Q^{\frac12}\,
                 Q^{-\frac12}\bigr\|_2^2\Bigr]\\
    &\le \zeta_{0,K}^2\,\big\|Q^{-\frac12}\big\|_2^2\,
      \Er\bigl[\big\|Q^{\frac12}\hat{Q}^{-1}Q^{\frac12}-I_K\big\|_2^2\bigr]\\
    &= \zeta_{0,K}^2\,\underline{\lambda}_K^{-1}\,
      \Er\bigl[\big\|Q^{\frac12}\hat{Q}^{-1}Q^{\frac12}-I_K\big\|_2^2\bigr].
  \end{aligned}
\]
Write \(M\coloneqq Q^{-\frac12}\hat{Q}Q^{-\frac12}\).  Then
\begin{align*}
    &\Er\bigl[\big\|Q^{\frac12}\hat{Q}^{-1}Q^{\frac12}-I_K\big\|_2^2\bigr]
    = \Er\bigl[\|M^{-1}-I_K\|_2^2\bigr]= \Er\bigl[\big\|M^{-\frac12}(M-I_K)M^{-\frac12}\big\|_2^2\bigr]\\
 \le& \Er\bigl[\|M^{-1}\|_2^2\,\|M-I_K\|_2^2\bigr]\le \|Q\|_2^2\,
           \Er\bigl[\|\hat{Q}^{-1}\|_2^2\,\|M-I_K\|_2^2\bigr] \le \zeta_{0,K}^4\,
           \Er\bigl[\|\hat{Q}^{-1}\|_2^2\,\|M-I_K\|_2^2\bigr].
\end{align*}
We bound the second piece by
\[
  \begin{aligned}
    \Er\Bigl[\big\|Q^{-\frac12}\bigl(P^\top \Psi_t/n
                         - \Er[p_K(\X)\psi_t(\X)]\bigr)\big\|^2\Bigr]&\le \Er\Bigl[\underline{\lambda}_K^{-1}
                    \big\|P^\top \Psi_t/n
                          - \Er[p_K(\X)\psi_t(\X)]\big\|^2\Bigr]\\
    & =\Er\Big[\underline{\lambda}_K^{-1}\Big\|\frac{1}{n}\sum_{i=1}^n \big(p_K(\X_i)\psi_t(\X_i)-\Er[p_K(\X)\psi_t(\X)]\big)\Big\|^2\Big]\\                      
    &= n^{-1}\,\underline{\lambda}_K^{-1}\,
           \Er\Bigl[\big\|p_K(\X)\,\psi_t(\X)
                        - \Er[p_K(\X)\psi_t(\X)]\big\|^2\Bigr]\\
    &\le \zeta_{0,K}^2\,n^{-1}\,\underline{\lambda}_K^{-1}.
  \end{aligned}
\]
Combining all the above bounds gives the result.
\end{proof}

\begin{proof}[Proof of \cref{lem:hatQ}]
Write $A_n\coloneqq \{\|\hat{Q}^{-1}\|_2^a\leq 2^a\underline{\lambda}_K^{-a}\}$. Note that 
\[
\begin{aligned}                 
  \Er[\|\hat{Q}^{-1}\|_2^a]&=\Er[\|\hat{Q}^{-1}\|_2^a\Ibbm(A_n)]+\Er[\|\hat{Q}^{-1}\|_2^a\Ibbm(A_n^c)]\leq 2^a\underline{\lambda}_K^{-a}+\lambda_n^{-a}\Prb(A_n^c),
\end{aligned}
\]
since $\|\hat{Q}^{-1}\|_2^a\leq \lambda_n^{-a}$ and $\lambda_n>0$. Therefore, it is sufficient to prove $\lambda_n^{-a}\Prb(A_n^c) \to 0$ as $n\to \infty$.

Define $\Delta\coloneqq \hat{Q}-Q$. The strategy is to first show $A_n^c\subseteq \{\|\Delta\|\geq \frac{1}{2}\underline{\lambda}_K\}$ and then apply the concentration inequality of the random matrix to bound $\Prb(\|\Delta\|\geq \frac{1}{2}\underline{\lambda}_K)$.

\Step[step1:setinclusion]{Show sets inclusion}
Now we show $A_n^c\subseteq \{\|\Delta\|\geq \frac{1}{2}\underline{\lambda}_K\}$. For that purpose,  suppose that $\|\Delta\|_2< \frac{1}{2}\underline{\lambda}_K$. This implies 
\[
\|Q^{-1}\Delta\|_2\leq \|Q^{-1}\|_2\|\Delta\|_2< \frac{1}{2},
\]
since $\|Q^{-1}\|_2=\underline{\lambda}_K^{-1}$. Then from the Neumann series representation of $(I_K+Q^{-1}\Delta)^{-1}$, we have 
\[
    (Q+\Delta)^{-1}=[Q(I_K+Q^{-1}\Delta)]^{-1}=\sum_{k=0}^\infty (Q^{-1}\Delta)^kQ^{-1}.
\]
Therefore,
\[
\|\hat{Q}^{-1}\|_2=\|(Q+\Delta)^{-1}\|_2\leq \|Q^{-1}\|_2\sum_{k=0}^\infty \|Q^{-1}\Delta\|_2^k=\frac{\|Q^{-1}\|_2}{1-\|Q^{-1}\Delta\|_2}\leq 2\underline{\lambda}_K^{-1}.
\]
This means that $\{\|\Delta\|_2< \frac{1}{2}\underline{\lambda}_K\}\subseteq A_n$, which is equivalent to $A_n^c \subseteq \{\|\Delta\|_2\geq \frac{1}{2}\underline{\lambda}_K\}$.

\Step[step2:bound]{Bound matrix perturbation}
Recall that 
\[
\Delta\coloneqq \hat{Q}-Q=\frac{1}{n}\sum_{i=1}^n(p_K(\X_i)p_K(\X_i)^\top -Q)+\lambda_n I_K,
\]
where the term $M_i\coloneqq \frac{1}{n}(p_K(\X_i)p_K(\X_i)^\top -Q)$ is mean-zero symmetric. From \cref{step1:setinclusion}, the concentration inequality of the random matrix \citep{tropp2012user} gives that (where we assume WLOG that $\underline{\lambda}_K-\lambda_n\geq 0$)
\[
\lambda_n^{-a}\Prb(A_n^c)\leq \lambda_n^{-a}\Prb\big(\|\Delta\|_2\geq \frac{1}{2}\underline{\lambda}_K\big)\leq \lambda_n^{-a}K\exp\Big(\frac{-(\frac{1}{2}\underline{\lambda}_K-\lambda_n)^2/2}{\sigma^2+R(\frac{1}{2}\underline{\lambda}_K-\lambda_n)/3}\Big),
\]
where $R$ is such that $\|M_i\|_2\leq R$ \as and $\sigma^2=\|\sum_{i=1}^n\Er[M_i^2]\|_2$. We can pick $R\coloneqq \frac{2\zeta_{0,K}^2}{n}$, since 
\[
    \|M_i\|_2\leq \frac{1}{n}(\|p_K(\X_i)p_K(\X_i)^\top\|_2+\|Q\|_2)\leq \frac{2\zeta_{0,K}^2}{n}. 
\]
We also have 
\[
\sigma^2=n\|\Er[M_1^2]\|_2\leq  4\zeta_{0,K}^4/n.
\]
Therefore, we have
\[
    \lambda_n^{-a}\Prb(A_n^c)\leq \lambda_n^{-a}K\exp\Big(\frac{-3n(\frac{1}{2}\underline{\lambda}_K-\lambda_n)^2}{32\zeta_{0,K}^4}\Big)+\lambda_n^{-a}K\exp\Big(\frac{-3n(\frac{1}{2}\underline{\lambda}_K-\lambda_n)}{16\zeta_{0,K}^2}\Big), 
\]
where the right-hand side goes to zero as $n\to \infty$ since $\lambda_n\asymp n^{-c}$ for some $c\geq 0$ and $\zeta_{0,K}=o\Big(\lb\frac{n(\underline{\lambda}_K-\lambda_n)^2}{\log K+\log(\lambda_n^{-a})}\rb^{\frac{1}{4}}\wedge \lb\frac{n(\underline{\lambda}_K-\lambda_n)}{\log K+\log(\lambda_n^{-a})}\rb^{\frac{1}{2}}\Big)$. This then concludes $\Er[\|\hat{Q}^{-1}\|_2^a]=O(\underline{\lambda}_K^{-a})$ as $n\to \infty$.
\end{proof}

\begin{proof}[Proof of \cref{lem:hatQ2}]
    
The Cauchy–Schwarz inequality gives
    \[
        \Er\big[\big\|\hat{Q}^{-1}\big\|_2^2\big\|Q^{-\frac{1}{2}}\hat{Q}Q^{-\frac{1}{2}}-I_K\big\|_2^2\big]\leq \Er\big[\big\|\hat{Q}^{-1}\big\|_2^4\big]^{\frac{1}{2}}\Er\big[\big\|Q^{-\frac{1}{2}}\hat{Q}Q^{-\frac{1}{2}}-I_K\big\|_2^4\big]^{\frac{1}{2}}.
    \]
    By \cref{lem:hatQ}, we have $\Er\big[\|\hat{Q}^{-1}\|_2^4\big]^{\frac{1}{2}}=O(\underline{\lambda}_K^{-2})$. So it remains to analyze $\Er\big[\big\|Q^{-\frac{1}{2}}\hat{Q}Q^{-\frac{1}{2}}-I_K\big\|_2^4\big]$.

    Set $\tilde{Q}\coloneqq \hat{Q}-\lambda_nI_K$. Then
    \[
    \Er\big[\big\|Q^{-\frac{1}{2}}\hat{Q}Q^{-\frac{1}{2}}-I_K\big\|_2^4\big]\leq 8\Er\big[\big\|Q^{-\frac{1}{2}}\tilde{Q}Q^{-\frac{1}{2}}-I_K\big\|_2^4\big]+8(\lambda_n\underline{\lambda}_K^{-1})^4.
    \]
  Define $\tilde{M}_i\coloneqq \frac{1}{n}(Q^{-\frac{1}{2}}p_K(\X)p_K(\X)^\top Q^{-\frac{1}{2}} -I_K)$. Then we have
    \[  
    \begin{aligned}
        \tilde{\sigma}^2&=n\big\|\Er\big[\tilde{M}_1^2\big]\big\|_2 \\
        &=\frac{1}{n}\big\|\Er\big[Q^{-\frac{1}{2}}p_K(\X)p_K(\X)^\top Q^{-\frac{1}{2}}-2Q^{-\frac{1}{2}}p_K(\X)p_K(\X)^\top Q^{-\frac{1}{2}}+I_K\big]\big\|_2\\
        &\leq \frac{1}{n}( 1+\underline{\lambda}_K^{-1}\zeta_{0,K}^2).
    \end{aligned}
    \]
    We have
    \[
        \big\|\tilde{M}_i\big\|_2\leq \frac{1}{n}\big\|Q^{-\frac{1}{2}}p_K(\X)p_K(\X)^\top Q^{-\frac{1}{2}}-I_K\big\|_2\leq \frac{1}{n}(\underline{\lambda}_K^{-1}\zeta_{0,K}^2+1)=: \tilde{R}.
    \]
Set $a_n\coloneqq \underline{\lambda}_K^{-\frac{1}{2}}\zeta_{0,K}\sqrt{\log(K)/n}+\underline{\lambda}_K^{-1}\zeta_{0,K}^2\log(K)/n=: b_n+c_n$. Applying the concentration inequality of the random matrix \citep{tropp2012user} gives
\[
\begin{aligned}
    \Er\big[\big\|Q^{-\frac{1}{2}}\tilde{Q}Q^{-\frac{1}{2}}&-I_K\big\|_2^4\big]=4\int_{0}^{\infty} t^{3}\Prb\big(\big\|Q^{-\frac{1}{2}}\tilde{Q}Q^{-\frac{1}{2}}-I_K\big\|_2\geq t\big) \dr t\\
    &=4\int_{0}^{a_n} t^{3}\Prb\big(\big\|Q^{-\frac{1}{2}}\tilde{Q}Q^{-\frac{1}{2}}-I_K\big\|_2\geq t\big) \dr t+4\int_{a_n}^{\infty} t^{3}\Prb\big(\big\|Q^{-\frac{1}{2}}\tilde{Q}Q^{-\frac{1}{2}}-I_K\big\|_2\geq t\big) \dr t\\
    &\leq a_n^4+4\int_{a_n}^\infty t^3K\left[\exp\Big(\frac{-3t^2}{8\tilde{\sigma}^2}\Big)\Ibbm\Big(t\leq \frac{\tilde{\sigma}^2}{\tilde{R}}\Big)+\exp\Big(\frac{-3t}{8\tilde{R}}\Big)\Ibbm\Big(t>\frac{\tilde{\sigma}^2}{\tilde{R}}\Big)\right]\dr t\\
    &\leq a_n^4+4\int_{b_n}^\infty t^3K\exp\Big(\frac{-t^2}{C\underline{\lambda}_K^{-1}n^{-1}\zeta_{0,K}^2}\Big)\dr t +4\int_{c_n}^\infty t^3K\exp\Big(\frac{-t}{C\underline{\lambda}_K^{-1}n^{-1}\zeta_{0,K}^2}\Big)\dr t\\
    &=: a_n^4+4\Upsilon_1+4\Upsilon_2.
\end{aligned}
\]

We now analyze $\Upsilon_1$:
\[
    \begin{aligned}
        \Upsilon_1&=\int_{b_n}^\infty t^3K\exp\Big(\frac{-t^2}{C\underline{\lambda}_K^{-1}n^{-1}\zeta_{0,K}^2}\Big)\dr t\\
        &=\frac{A^2K}{2}\int_{\frac{b_n^2}{A}}^\infty u\exp(-u)\dr u & (A\coloneqq C\underline{\lambda}_K^{-1}n^{-1}\zeta_{0,K}^2,u\coloneqq \frac{t^2}{A})\\
        &=\frac{A^2K}{2}\Big(1+\frac{b_n^2}{A}\Big)\exp\Big(-\frac{b_n^2}{A}\Big)  & (\int_x^\infty u\er^{-u}\dr u=(1+x)\er^{-x})\\
        &=O\big(\underline{\lambda}_K^{-2}n^{-2}\zeta_{0,K}^4+\underline{\lambda}_K^{-2}n^{-2}\zeta_{0,K}^4\log K\big) & \text{ as } n\to \infty.
    \end{aligned}
\]
Since $\underline{\lambda}_K^{-1}\zeta_{0,K}^2\log(K)n^{-1} \to 0 $ as $n\to \infty$, we have $a_n\asymp \underline{\lambda}_K^{-\frac{1}{2}}\zeta_{0,K}\sqrt{\log(K)/n}$ and therefore $a_n^4\asymp \underline{\lambda}_K^{-2}\zeta_{0,K}^4(\log K)^2n^{-2}$. Also, recall that $K\to \infty$ as $n\to \infty$. Hence, 
\[
    \frac{\underline{\lambda}_K^{-2}n^{-2}\zeta_{0,K}^4+\underline{\lambda}_K^{-2}n^{-2}\zeta_{0,K}^4\log K}{a_n^4} \to 0\quad \text{ as } n\to \infty,
\]
which implies $\Upsilon_1=o(a_n^4)$.

Next, we analyze $\Upsilon_2$:
\[
    \begin{aligned}
        \Upsilon_2&=\int_{c_n}^\infty t^3K\exp\Big(\frac{-t}{C\underline{\lambda}_K^{-1}n^{-1}\zeta_{0,K}^2}\Big)\dr t\\
        &= KA^4\Gamma\big(4,\frac{c_n}{A}\big) & (A\coloneqq C\underline{\lambda}_K^{-1}n^{-1}\zeta_{0,K}^2)\\
        &=3!KA^4\exp\Big(\frac{-c_n}{A}\Big)\Big(1+\frac{c_n}{A}+\frac{c_n^2}{2A^2}+\frac{c_n^3}{6A^3})\qquad  & (s\in\Zb^+, \Gamma(s,x)=(s-1)!\er^{-x}\sum_{k=0}^{s-1}\frac{x^k}{k!}\Big)\\
        &=O\big(\underline{\lambda}_K^{-4}n^{-4}\zeta_{0,K}^8(\log K)^3\big).
    \end{aligned}
\]  
Since $\underline{\lambda}_K^{-1}\zeta_{0,K}^2\log(K)n^{-1} \to 0 $ as $n\to \infty$, it holds
\[
\frac{\underline{\lambda}_K^{-4}n^{-4}\zeta_{0,K}^8(\log K)^3}{\underline{\lambda}_K^{-2}\zeta_{0,K}^4(\log K)^2n^{-2}}=\underline{\lambda}_K^{-2}n^{-2}\zeta_{0,K}^4 \log K\to 0, \quad\text{ as }n \to \infty.
\]
Hence, $\Upsilon_2=o(a_n^4)$. Overall, it establishes 
\[
    \Er[\|Q^{-\frac{1}{2}}\hat{Q}Q^{-\frac{1}{2}}-I_K\|_2^4]=O(a_n^4+(\lambda_n\underline{\lambda}_K^{-1})^4)=O(\underline{\lambda}_K^{-2}\zeta_{0,K}^4\log(K)^2n^{-2}+(\lambda_n\underline{\lambda}_K^{-1})^4)
\]
and thus finishes the proof.
\end{proof}

{\small
\bibliographystyle{apalike}
\bibliography{AMS}

@article{zhao2022analysis,
	Author = {Zhao, Puning and Lai, Lifeng},
	Journal = {IEEE Transactions on Information Theory},
	Number = {12},
	Pages = {7971--7995},
	Publisher = {IEEE},
	Title = {Analysis of k{NN} density estimation},
	Volume = {68},
	Year = {2022}}

@article{singh2016finite,
	Author = {Singh, Shashank and P{\'o}czos, Barnab{\'a}s},
	Journal = {Advances in Neural Information Processing Systems},
	Title = {Finite-sample analysis of fixed-k nearest neighbor density functional estimators},
	Volume = {29},
	Year = {2016}}

@book{delfour2011shapes,
	Author = {Delfour, M.C. and Zolesio, J.P.},
	Isbn = {9780898719369},
	Lccn = {2010028846},
	Publisher = {Society for Industrial and Applied Mathematics},
	Series = {Advances in Design and Control},
	Title = {Shapes and Geometries: Metrics, Analysis, Differential Calculus, and Optimization, Second Edition},
	Url = {https://books.google.nl/books?id=V773JJZd6RMC},
	Year = {2011}}

@book{Grisvard11Elliptic_nonsmooth_domains,
	Author = {Grisvard, Pierre},
	Doi = {10.1137/1.9781611972030},
	Eprint = {https://epubs.siam.org/doi/pdf/10.1137/1.9781611972030},
	Publisher = {Society for Industrial and Applied Mathematics},
	Title = {Elliptic Problems in Nonsmooth Domains},
	Url = {https://epubs.siam.org/doi/abs/10.1137/1.9781611972030},
	Year = {2011}}

@article{viel2025convergenceratenearestneighbour,
	Author = {Simon Viel and Lionel Truquet and Ikko Yamane},
	Journal = {arXiv preprint arXiv:2504.21633},
	Title = {Convergence rate for Nearest Neighbour matching: geometry of the domain and higher-order regularity},
	Url = {https://arxiv.org/abs/2504.21633},
	Year = {2025}}

@book{evans2015measure,
	Author = {Evans, L.C. and Gariepy, R.F.},
	Isbn = {9781482242393},
	Publisher = {CRC Press},
	Series = {Textbooks in Mathematics},
	Title = {Measure Theory and Fine Properties of Functions, Revised Edition},
	Url = {https://books.google.nl/books?id=e3R3CAAAQBAJ},
	Year = {2015}}

@article{ansari2025directextensionazadkia,
	Author = {Ansari, Jonathan and Fuchs, Sebastian},
	Journal = {arXiv preprint arXiv:2212.01621},
	Title = {A direct extension of {A}zadkia \& {C}hatterjee's rank correlation to multi-response vectors},
	Year = {2025}}

@article{evans2002asymptotic,
	Author = {Evans, Dafydd and Jones, Antonia J and Schmidt, Wolfgang M},
	Journal = {Proceedings of the Royal Society of London. Series A: Mathematical, Physical and Engineering Sciences},
	Number = {2028},
	Pages = {2839--2849},
	Publisher = {The Royal Society},
	Title = {Asymptotic moments of near--neighbour distance distributions},
	Volume = {458},
	Year = {2002}}

@article{kurisu2024series,
	Author = {Kurisu, Daisuke and Matsuda, Yasumasa},
	Journal = {arXiv preprint arXiv:2402.02773},
	Title = {Series ridge regression for spatial data on $\mathbb{R}^d$},
	Year = {2024}}

@article{tuo2024asymptotic,
	Author = {Tuo, Rui and Zou, Lu},
	Journal = {arXiv preprint arXiv:2403.04248},
	Title = {Asymptotic Theory for Linear Functionals of Kernel Ridge Regression},
	Year = {2024}}

@incollection{chatterjee2022survey,
	Author = {Chatterjee, Sourav},
	Booktitle = {Probability and Stochastic Processes, A Volume in Honour of Rajeeva L. Karandikar},
	Journal = {arXiv preprint arXiv:2211.04702},
	Publisher = {Springer},
	Title = {A survey of some recent developments in measures of association},
	Year = {2024}}

@article{zhou2025association,
	Author = {Zhou, Hang and M{\"u}ller, Hans-Georg},
	Journal = {arXiv preprint arXiv:2505.01983},
	Title = {Association and Independence Test for Random Objects},
	Year = {2025}}

@article{zhang2025extensions,
	Author = {Zhang, Qingyang},
	Journal = {Journal of Nonparametric Statistics},
	Pages = {1--30},
	Publisher = {Taylor \& Francis},
	Title = {On the extensions of the {C}hatterjee-{S}pearman test},
	Year = {2025}}

@article{zhang2024asymptotic,
	Author = {Zhang, Qingyang},
	Journal = {arXiv preprint arXiv:2401.05281},
	Title = {Asymptotic expected sensitivity function and its applications to nonparametric correlation estimators},
	Year = {2024}}

@article{gao2024family,
	Author = {Gao, Muhong and Li, Qizhai},
	Journal = {arXiv preprint arXiv:2403.17670},
	Title = {A family of {C}hatterjee's correlation coefficients and their properties},
	Year = {2024}}

@article{lin2024failure,
	Author = {Lin, Zhexiao and Han, Fang},
	Journal = {Biometrika},
	Number = {3},
	Pages = {1063--1070},
	Publisher = {Oxford University Press},
	Title = {On the failure of the bootstrap for {C}hatterjee's rank correlation},
	Volume = {111},
	Year = {2024}}

@article{bickel2022measures,
	Author = {Bickel, Peter J},
	Journal = {arXiv preprint arXiv:2206.13663},
	Title = {Measures of independence and functional dependence},
	Year = {2022}}

@article{olivares2025powerful,
	Author = {Olivares, Mauricio and Olma, Tomasz and Wilhelm, Daniel},
	Journal = {arXiv preprint arXiv:2503.21715},
	Title = {A Powerful Bootstrap Test of Independence in High Dimensions},
	Year = {2025}}

@article{ansari2023quantifying,
	Author = {Ansari, Jonathan and Langthaler, Patrick B and Fuchs, Sebastian and Trutschnig, Wolfgang},
	Journal = {arXiv preprint arXiv:2308.06168},
	Title = {Quantifying and estimating dependence via sensitivity of conditional distributions},
	Year = {2023}}

@article{han2024azadkia,
	Author = {Han, Fang and Huang, Zhihan},
	Journal = {The Annals of Applied Probability},
	Number = {6},
	Pages = {5172--5210},
	Publisher = {Institute of Mathematical Statistics},
	Title = {Azadkia--{C}hatterjee's correlation coefficient adapts to manifold data},
	Volume = {34},
	Year = {2024}}

@article{tran2024rank,
	Author = {Tran, Leon and Han, Fang},
	Journal = {arXiv preprint arXiv:2412.02668},
	Title = {On a rank-based {A}zadkia-{C}hatterjee correlation coefficient},
	Year = {2024}}

@article{azadkia2025new,
	Author = {Azadkia, Mona and Roudaki, Pouya},
	Journal = {arXiv preprint arXiv:2505.18146},
	Title = {A new measure of dependence: Integrated $R^2$},
	Year = {2025}}

@article{azadkia2021fast,
	Author = {Azadkia, Mona and Taeb, Armeen and B{\"u}hlmann, Peter},
	Journal = {arXiv preprint arXiv:2111.14969},
	Title = {A fast non-parametric approach for causal structure learning in polytrees},
	Year = {2021}}

@article{gamboa2022global,
	Author = {Gamboa, Fabrice and Gremaud, Pierre and Klein, Thierry and Lagnoux, Agn{\`e}s},
	Journal = {Bernoulli},
	Number = {4},
	Pages = {2345--2374},
	Publisher = {Bernoulli Society for Mathematical Statistics and Probability},
	Title = {Global sensitivity analysis: A novel generation of mighty estimators based on rank statistics},
	Volume = {28},
	Year = {2022}}

@article{fuchs2024quantifying,
	Author = {Fuchs, Sebastian},
	Journal = {Journal of Multivariate Analysis},
	Pages = {105266},
	Publisher = {Elsevier},
	Title = {Quantifying directed dependence via dimension reduction},
	Volume = {201},
	Year = {2024}}

@article{griessenberger2022multivariate,
	Author = {Griessenberger, Florian and Junker, Robert R and Trutschnig, Wolfgang},
	Journal = {Electronic Journal of Statistics},
	Number = {1},
	Pages = {2206--2251},
	Publisher = {The Institute of Mathematical Statistics and the Bernoulli Society},
	Title = {On a multivariate copula-based dependence measure and its estimation},
	Volume = {16},
	Year = {2022}}

@article{bucher2024lack,
	Author = {B{\"u}cher, Axel and Dette, Holger},
	Journal = {arXiv preprint arXiv:2410.11418},
	Title = {On the lack of weak continuity of {C}hatterjee's correlation coefficient},
	Year = {2024}}

@article{strothmann2024rearranged,
	Author = {Strothmann, Christopher and Dette, Holger and Siburg, Karl Friedrich},
	Journal = {Bernoulli},
	Number = {2},
	Pages = {1055--1078},
	Publisher = {Bernoulli Society for Mathematical Statistics and Probability},
	Title = {Rearranged dependence measures},
	Volume = {30},
	Year = {2024}}

@article{dette2025simple,
	Author = {Dette, Holger and Kroll, Marius},
	Journal = {Biometrika},
	Number = {1},
	Pages = {asae045},
	Publisher = {Oxford University Press},
	Title = {A simple bootstrap for {C}hatterjee's rank correlation},
	Volume = {112},
	Year = {2025}}

@article{kroll2024asymptotic,
	Author = {Kroll, Marius},
	Journal = {arXiv preprint arXiv:2408.11547},
	Title = {Asymptotic Normality of {C}hatterjee's Rank Correlation},
	Year = {2024}}

@article{cattaneo2024rosenbaum,
	Author = {Cattaneo, Matias D and Han, Fang and Lin, Zhexiao},
	Journal = {Biometrika},
	Number = {1},
	Pages = {asae062},
	Publisher = {Oxford University Press},
	Title = {On {R}osenbaum's rank-based matching estimator},
	Volume = {112},
	Year = {2025}}

@article{Azadkia21simple,
	Author = {Mona Azadkia and Sourav Chatterjee},
	Doi = {10.1214/21-AOS2073},
	Journal = {The Annals of Statistics},
	Number = {6},
	Pages = {3070--3102},
	Publisher = {Institute of Mathematical Statistics},
	Title = {{A simple measure of conditional dependence}},
	Url = {https://doi.org/10.1214/21-AOS2073},
	Volume = {49},
	Year = {2021}}

@article{lin2022limit,
	Author = {Zhexiao Lin and Fang Han},
	Journal = {arXiv.org preprint},
	Title = {Limit theorems of {C}hatterjee's rank correlation},
	Url = {https://arxiv.org/abs/2204.08031},
	Volum = {arXiv:2204.08031 [math.ST]},
	Year = {2022}}

@article{tropp2012user,
	Author = {Tropp, Joel A},
	Journal = {Foundations of Computational Mathematics},
	Pages = {389--434},
	Publisher = {Springer},
	Title = {User-friendly tail bounds for sums of random matrices},
	Volume = {12},
	Year = {2012}}

@article{belloni2015some,
	Author = {Belloni, Alexandre and Chernozhukov, Victor and Chetverikov, Denis and Kato, Kengo},
	Journal = {Journal of Econometrics},
	Number = {2},
	Pages = {345--366},
	Publisher = {Elsevier},
	Title = {Some new asymptotic theory for least squares series: Pointwise and uniform results},
	Volume = {186},
	Year = {2015}}

@article{lin2022regression,
	Author = {Lin, Zhexiao and Han, Fang},
	Journal = {Journal of Econometrics},
	Pages = {106080},
	Title = {On regression-adjusted imputation estimators of average treatment effects},
	Volume = {251},
	Year = {2025}}

@article{chen2018optimal,
	Author = {Chen, Xiaohong and Christensen, Timothy M},
	Journal = {Quantitative Economics},
	Number = {1},
	Pages = {39--84},
	Publisher = {Wiley Online Library},
	Title = {Optimal sup-norm rates and uniform inference on nonlinear functionals of nonparametric {IV} regression},
	Volume = {9},
	Year = {2018}}

@article{rubin1973use,
	Author = {Rubin, Donald B},
	Journal = {Biometrics},
	Number = {1},
	Pages = {185--203},
	Publisher = {JSTOR},
	Title = {The use of matched sampling and regression adjustment to remove bias in observational studies},
	Volume = {29},
	Year = {1973}}

@article{lin2021estimation,
	Author = {Lin, Zhexiao and Ding, Peng and Han, Fang},
	Journal = {Econometrica},
	Number = {6},
	Pages = {2187--2217},
	Title = {Estimation based on nearest neighbor matching: from density ratio to average treatment effect},
	Volume = {91},
	Year = {2023}}

@article{han2021extensions,
	Author = {Han, Fang},
	Journal = {Bernoulli News},
	Pages = {7--11},
	Title = {On extensions of rank correlation coefficients to multivariate spaces},
	Volume = {28},
	Year = {2021}}

@article{shi2020power,
	Author = {Shi, H and Drton, M and Han, F},
	Doi = {10.1093/biomet/asab028},
	Eprint = {https://academic.oup.com/biomet/advance-article-pdf/doi/10.1093/biomet/asab028/40545355/asab028.pdf},
	Issn = {0006-3444},
	Journal = {Biometrika},
	Number = {2},
	Pages = {317--333},
	Title = {{On the power of Chatterjee's rank correlation}},
	Url = {https://doi.org/10.1093/biomet/asab028},
	Volume = {109},
	Year = {2022}}

@article{shi2021ac,
	Author = {Shi, Hongjian and Drton, Mathias and Han, Fang},
	Journal = {Bernoulli},
	Number = {2},
	Pages = {851--877},
	Title = {On {A}zadkia-{C}hatterjee's conditional dependence coefficient},
	Volume = {30},
	Year = {2024}}

@article{newey1997convergence,
	Author = {Newey, Whitney K},
	Journal = {Journal of Econometrics},
	Number = {1},
	Pages = {147--168},
	Publisher = {Elsevier},
	Title = {Convergence rates and asymptotic normality for series estimators},
	Volume = {79},
	Year = {1997}}

@article{abadie2011bias,
	Author = {Abadie, A. and Imbens, G. W.},
	Journal = {Journal of Business and Economic Statistics},
	Number = {1},
	Pages = {1--11},
	Title = {Bias-corrected matching estimators for average treatment effects},
	Volume = {29},
	Year = {2011}}

@article{auddy2021exact,
	Author = {Auddy, Arnab and Deb, Nabarun and Nandy, Sagnik},
	Journal = {Bernoulli},
	Number = {2},
	Pages = {1640--1668},
	Title = {Exact detection thresholds and minimax optimality of {C}hatterjee's correlation coefficient},
	Volume = {30},
	Year = {2024}}

@article{MR3909934,
	Author = {Berrett, Thomas B. and Samworth, Richard J. and Yuan, Ming},
	Doi = {10.1214/18-AOS1688},
	Fjournal = {The Annals of Statistics},
	Issn = {0090-5364},
	Journal = {The Annals of Statistics},
	Mrclass = {62G05 (62G20)},
	Mrnumber = {3909934},
	Number = {1},
	Pages = {288--318},
	Title = {Efficient multivariate entropy estimation via {$k$}-nearest neighbour distances},
	Url = {https://doi.org/10.1214/18-AOS1688},
	Volume = {47},
	Year = {2019}}

@article{chatterjee2020new,
	Author = {Chatterjee, Sourav},
	Doi = {10.1080/01621459.2020.1758115},
	Fjournal = {Journal of the American Statistical Association},
	Issn = {0162-1459},
	Journal = {Journal of the American Statistical Association},
	Number = {535},
	Pages = {2009--2022},
	Title = {A new coefficient of correlation},
	Url = {https://doi.org/10.1080/01621459.2020.1758115},
	Volume = {116},
	Year = {2021}}

@unpublished{deb2020kernel,
	Author = {Deb, Nabarun and Ghosal, Promit and Sen, Bodhisattva},
	Note = {Available at \href{https://arxiv.org/abs/2010.01768v2}{arXiv:2010.01768v2}},
	Title = {Measuring association on topological spaces using kernels and geometric graphs},
	Year = {2020}}

@article{MR3024030,
	Author = {Dette, Holger and Siburg, Karl F. and Stoimenov, Pavel A.},
	Doi = {10.1111/j.1467-9469.2011.00767.x},
	Fjournal = {Scandinavian Journal of Statistics. Theory and Applications},
	Issn = {0303-6898},
	Journal = {Scandinavian Journal of Statistics},
	Mrclass = {62H20 (60E15 62G08 62G20)},
	Mrnumber = {3024030},
	Number = {1},
	Pages = {21--41},
	Title = {A copula-based non-parametric measure of regression dependence},
	Url = {https://doi.org/10.1111/j.1467-9469.2011.00767.x},
	Volume = {40},
	Year = {2013}}

@article{huang2020kernel,
	Author = {Huang, Zhen and Deb, Nabarun and Sen, Bodhisattva},
	Journal = {Journal of Machine Learning Research},
	Note = {Available at \href{https://arxiv.org/abs/2012.14804v1}{arXiv:2012.14804v1}},
	Number = {216},
	Pages = {1--58},
	Title = {Kernel partial correlation coefficient -- a measure of conditional dependence},
	Volume = {23},
	Year = {2022}}
}

\end{document}